\documentclass[fleqn,usenatbib]{mnras}

\usepackage{newtxtext,newtxmath}

\usepackage[T1]{fontenc}

\DeclareRobustCommand{\VAN}[3]{#2}
\let\VANthebibliography\thebibliography
\def\thebibliography{\DeclareRobustCommand{\VAN}[3]{##3}\VANthebibliography}

\usepackage{graphicx}	
\usepackage{amsmath}	

\usepackage{amssymb}	

\title[Megamasers in MIR Spectra]{Searching Water Megamasers By Using Mid-infrared Spectroscopy (I): Possible Mid-infrared Indicators}

\author[Lam et al.]{Man I Lam$^{1,2}$\thanks{E-mail: mlam@aip.de, mlam.astro1912@gmail.com}, 
C. Jakob Walcher$^{1}$, 
Feng Gao$^{3}$, 
Ming Yang$^{4}$, 
Huan Li$^{5}$, 
Lei Hao$^{6}$ \\
$^{1}$ Leibniz-Institut für Astrophysik Potsdam (AIP), An der Sternwarte 16, 14482 Potsdam, Germany \\
$^{2}$ National Astronomical Observatories, CAS, 20A Datun Rd., Chao-Yang Dist., 100101, Beijing, PR China \\
$^{3}$ Max Planck Institute for Extraterrestrial Physics (MPE), Giessenbachstrasse 1, 85748 Garching, Germany \\
$^{4}$ IAASARS, National Observatory of Athens, Vas. Pavlou and I. Metaxa, 15236 Penteli, Greece \\
$^{5}$ School of Aerospace Science and Technology, Xidian University, 266 Xinglong Section of Xifeng Road, 710071 Xi'an, PR China\\
$^{6}$ Shanghai Astronomical Observatory, 80 Nandan Rd., 200030 Shanghai, PR China
}

\date{Accepted XXX. Received YYY; in original form ZZZ}

\pubyear{2020}

\begin{document}
\label{firstpage}
\pagerange{\pageref{firstpage}--\pageref{lastpage}}
\maketitle

\begin{abstract}
Water megamasers at 22 GHz with a gas disk configuration in galaxies provide the most precise measurements of supermassive black hole masses, as well as independent constraints on the Hubble constant in the nearby universe. The existence of other maser types, such as jet or outflow masers, represents another tracer for AGN science. However, the detection rate of water megamasers in galaxies is extremely low. Over 40 years, only $\sim$ 160 galaxies are found to harbour maser emission, and $\sim$30\% of them show features in their maser emission that indicate a disk-like geometry. Therefore, increasing the detection rate of masers is a crucial task to allow expanding on maser studies. We present a comparison of mid-infrared spectroscopic data between a maser galaxy sample and a Seyfert 2 control sample. We find that maser galaxies show significant peculiarities in their mid-infrared spectra: (1) Maser galaxies tend to present stronger silicate absorption at $\tau_{9.7 \mu m}$ than the control sample,  (2) PAH 11.3 $\mu$m emission in maser galaxies is much weaker than in the control sample, (3) spectral indices at 20 - 30 $\mu$m are steeper in maser galaxies than in the control sample and tend to be mid-infrared enhanced population. We conclude that there may be good indicators in mid-infrared and far-infrared which could differentiate maser and non-maser Seyfert 2 galaxies. Upcoming infrared facilities, such as the \textit{James Webb Space Telescope}, may be able to exploit these and other useful criteria and tracers for water megamaser observations. 
\end{abstract}




\section{Introduction}
Maser emission is a naturally amplified, monochromatic, stimulated emission that can be observed, at radio wavelengths, also in astrophysical sources outside our Galaxy (see Review by \citealt{Lo2005,Tarchi2012,Henkel2018}). Historically, based on the maser luminosity, extragalactic masers are classified as kilomasers ($\textrm{L}_{\textrm{masers}}< $ 10 L$_{\odot}$) and megamasers ($\textrm{L}_{\textrm{masers}}> $ 10 L$_{\odot}$). The most common found masers in extragalactic environment are OH and water masers. 
The water maser also frequently appeared in the innermost regions of galaxies, which makes it an interesting tracer for galactic processes, in particular, those associated with active galactic nuclei  (AGN; e.g. \citealt{Tarchi2012,Peck2003,Greenhill2003}). The 22 GHz water megamasers with disk configuration (`disk masers') often show edge-on, thin Keplerian disks on sub-parsec scales around the central super-massive black holes (SMBHs) and the enclosed mass can be constrained to $\pm$3\% \citep{Humphreys2013}. This provides the most precise way to measure the mass of SMBHs and geometric distances, as well as constrain the cosmological model and the nature of dark energy (e.g. \citealt{Greenhill1995, Miyoshi1995, Braatz2010, Kuo2011,Kuo2013, Reid2013,Gao2016,Pesce2020}). In addition, the existent of other masers types (e.g. jet/outflow), which are not associated to accretion disks, constitute another interesting group for AGN sciences (e.g. \citealt{Tarchi2012} and references therein; \citealt{Surcis2020}).

Although 22 GHz water megamasers (hereafter masers or water masers) are excellent candidates to tackle many issues in modern astrophysics, it is a huge challenge to increase the sample size of known masers. Over a decade, a lot of efforts have been made to increase the number of detected maser galaxies, which however, still remains low. The Mega-maser Cosmology Project (MCP) has searched over 6000 active galactic nuclei (AGN) in the nearby universe \citep{Braatz2010,Kuo2018, Kuo2019}, but only 3\% of them have been detected with masers. Among all maser galaxies, $\sim$30\% may originate in a disk derived mostly from single-dish spectra ($\sim$0.9\% of the entire MCP sample) (e.g. \citealt{Kuo2020}).  

So far, megamasers are mostly detected in narrow-line AGN, such as Seyfert 2 or low-ionization nuclear emission-line regions (LINERs; e.g. \citealt{Braatz1997, Kondratko2006a}). Some luminous masers have been, however, also detected in some other classes of galaxies, including Type 1 AGN (e.g. \citealt{Hagiwara2003,Tarchi2011}). It is also believed that the maser galaxies tend to have higher column densities of obscuring material in the X-rays ($N_{\textrm{H}}$) than non-maser galaxies (e.g.\citealt{Kondratko2006b,Greenhill2008,Zhang2010}). \citet{Kondratko2006b} found a weak correlation between maser luminosity and X-ray luminosity, indicating that X-ray luminosity might be a tracer to find maser galaxies. Besides, \citet{Castangia2016} and \citet{Castangia2019}, based on a selection criterion presented in \citet{Severgnini2012}, suggested that the combination of mid-infrared (MIR) and X-ray luminosity might be able to increase the detection rate of masers. \citet{Panessa2020} suggested the maser detection rate was significantly increased in Compton-thick AGNs and hard X-ray luminosity. Moreover, there are a few characteristic spectral features of maser galaxies which are different from other galaxies, and these may be helpful to increase the detection rate. For example, \citet{Zhu2011} \& \citet{Constantin2012} found that maser galaxies tend to show higher intrinsic [O$\textrm{\sevensize{III}}$] $\lambda$5007 luminosities and higher obscurations than non-maser galaxies. Moreover, based on the [S$\textrm{\sevensize{II}}$] doublet ratios, they also found narrow ranges in Eddington ratios and electron densities of maser galaxies. However, only a small fraction of maser galaxies are covered by SDSS or other spectroscopic surveys, which makes it difficult to study the sample in detail. In the longer wavelength range, there are also a few attempts to increase the maser detection rate. \citet{Zhang2012,Zhang2018} studied radio continuum emission of maser and non-maser galaxies, and they found maser galaxies tend to show higher radio continuum at 6 cm and 20 cm than non-maser galaxies. 

The infrared bands may contain potential sensitive maser indicators, given that the excitation temperatures of the water megamasers requires at a temperature of $>$ 300 K \citep{Kylafis1991}. Theoretical studies (e.g. \citealt{Yates1997,Maloney2002}) also suggest the co-existence of dust and H$_2$O molecules where maser emission occurs. 
The dust surrounding AGNs could produce strong MIR emission which is distinguishable between star-forming, unobscured AGNs, and obscured AGNs populations. It has been found that the MIR luminosity is useful to classify unobscured and obscured AGNs (e.g. \citealt{Lacy2004,Stern2012,Assef2013}). \citet{Kuo2018} suggested a higher average luminosity of W4 band (22 $\mu$m) for maser galaxies based on an all-sky survey by Wide-field Infrared Survey Explorer (\textit{WISE}; \citealt{Wright2010}), which might indicate that the detection rate can be increased if maser candidates were selected by using their 22 $\mu$m luminosities. The result also indicated that maser galaxies might be a dusty population. However, the sensitivity of W4 band is much lower compared to W1 and W2 bands in \textit{WISE}. Moreover, the W4 luminosity is largely contaminated by star formation regions in host galaxies, where the exact contribution from maser is difficult to be determined since W4 is a broad filter. Indeed, a maser in a Seyfert 2 may also be not associated with the nucleus of the galaxy and be produced by (off-)nuclear star-formation. Yet, studying the difference of MIR spectra between maser and control sample galaxies may be able to shed some light on differentiating the two populations. 

This paper is the first in a series reporting the water megamasers by selecting in mid-infrared spectroscopy. In Paper I, we aim at comparing the infrared properties of a maser sample with respect to a Seyfert 2 control sample, in particular, looking for a sensitive tracer to differentiate two populations. Our future studies will further explore the correlation between maser luminosity and high excitation lines (Paper II, Li et al, in prep.) and the dust temperature in maser host galaxies (Paper III, Lam et al., in prep.).  This paper is structured as follows. The infrared data description is presented in section 2. The maser and control sample selection are presented in section 3. Infrared spectral indices, silicate strength at 9.7 $\mu$m, as well as PAH feature measurements, are presented in section 4. The observational results are discussed in light of the possible maser emission mechanisms and of increasing detection rates in section 5. Finally, a summary is given in section 6.

\section{Mid-Infrared Data}

\subsection{Infrared Spectra}
The most comprehensive sample of MIR spectra are provided by \textit{Spitzer}/Infrared Spectrograph (IRS). The \textit{Spitzer}/IRS contains both low- and high resolution mode covering about 5 - 38 $\mu$m (each mode also has both short and long slits). In low resolution mode, the slit sizes of short-low mode are 3.7\arcsec $\times$ 57\arcsec~and 3.6\arcsec $\times$ 57 \arcsec, corresponding to a wavelength coverage of 7.4 - 14.5 $\mu$m and 5.2 - 7.7 $\mu$m, respectively. The slit sizes of the long-low mode are 10.7\arcsec $\times$ 168 \arcsec~and 10.5 \arcsec $\times$ 168 $\arcsec$, corresponding to a wavelength coverage of 19.5 - 38.0 $\mu$m and 14.0 - 21.3 $\mu$m, respectively. The spectral resolution of the low resolution mode is $R = \lambda / \delta \lambda \approx $ 60 in the wavelength ranges. In high resolution mode, the slit size of the short-high mode is 4.7\arcsec $\times$ 11.3\arcsec, corresponding to a wavelength coverage of 9.9 - 19.6 $\mu$m. The slit size of the long-high mode is 11.1\arcsec $\times$ 22.3\arcsec, corresponding to a wavelength coverage of 18.7 - 37.2 $\mu$m. The spectral resolution of the high resolution mode is $R = \lambda / \delta \lambda \approx $ 600 (More information listed in Table~\ref{tab:IRS_table}). 

\subsection{Infrared Photometry}


The \textit{WISE} all-sky survey contains hundreds of millions of objects at 3.4, 4.6, 12, 22$\mu$m (W1, W2, W3, W4) with spatial resolutions of 6.1\arcsec, 6.4\arcsec, 6.5\arcsec, 12.0\arcsec, respectively \citep{Wright2010,Yan2013}.
The \textit{ALLWISE} program is an extended project based on \textit{WISE}, which combines both cryogenic and post-cryogenic surveys and covers the entire sky with about 700 million detected targets. We adopted the same cross-matching radius of 6.0\arcsec as had been done in \citet{Lambrides2019} \textbf{(see below)}. We obtained 4 bands photometric data of all maser and control galaxies in this study. The photometric data of maser galaxies is listed in full Table~\ref{tab:masers_table}.

\section{Masers and Control Sample With Infrared Spectroscopy}

Hereafter, we will refer to the confirmed H$_2$O maser emission objects `maser galaxies' or `masers', and galaxies used for comparison as `control galaxies', or `control sample'.

\subsection{Sample of Water MegaMaser Galaxies}
\label{sec:masers} 
The latest and most comprehensive sample of galaxies observed in a search of water maser emission is collected by the MCP \citep{Braatz2010,Braatz2018}, which contains $\geq$ 6,000 galaxies in the redshift range of z $\sim$ 0.0 - 0.07 \citep{Kuo2019}. However, the detection rate of maser galaxies is low. Only $\sim$3\% of observed galaxies have a detected H$_2$O maser as listed on the MCP website\footnote{https://safe.nrao.edu/wiki/bin/view/Main/MegamaserCosmologyProject/}. The list includes sky positions, recession velocities, maser luminosities, maser morphological classifications (e.g. both confirmed and putative disk, jet masers), and the corresponding reference for each detection. Besides, the observed maser spectra, the sensitivity of the observations and the corresponding source brightness temperature are also reported. The maser luminosities are calculated under the assumption of isotropic emission \citep{Pesce2015}. \citet{Kuo2018} listed 161 low-$z$ confirmed maser galaxies from MCP, which only included galaxies at lower redshift (D $<$ 180 Mpc). Among them, 34 galaxies ($\sim$21\%) are classified as disk masers. We adopted the total maser sample of 161 targets from \citet{Kuo2018} with classification. In order to get a complete maser type classification, we also cross-matched maser catalogue from \citet{Panessa2020} and replenish the additional maser types (e.g. jet/outflow). 

\citet{Lambrides2019} used `web scraping' to extract observational information of extragalactic objects with SMBHs and star-formation from \textit{Spitzer} Heritage Archive (SHA), and derived their emission line fluxes based on \textit{Spitzer}/IRS spectra (notice that they only took into account the low resolution mode of IRS spectra with a signal-to-noise ratio (SNR) larger than three. The sample consists of 2,015 galaxies occupying a wide range of infrared luminosity ($\nu L_{[22 \mu m]} = 10^8 - 10^{12} L_{\odot}$). The catalogue also provides additional information such as sky positions, redshift, photometry from \textit{WISE} and 2MASS, luminosities of PAH features at 6.6 $\mu$m, 7.7 $\mu$m and 11.3 $\mu$m, as well as H$_2$ rotational-vibrational emissions S(0) - S(7). 

We cross-matched the water maser sample of MCP and MIR spectral sample of \citet{Lambrides2019} by using a search radius of 3$\arcsec$ to retrieve maser galaxies with MIR spectral measurements, which resulted in a final sample of 43 MIR maser galaxies as listed in Table~\ref{tab:masers_table} (the full table is available online). We collected MIR spectra of the final sample from \textit{Spitzer} Enhanced Imaging Products (SEIP), which included super mosaic images, spectra and a source list for objects observed by \textit{Spitzer} (the same AORkeys used in \citealt{Lambrides2019} were adopted to retrieve the spectra).

The final sample of 43 masers with Spitzer MIR spectra, which includes 8 confirmed disk masers, 4 jet/outflow masers, and 1 star-forming maser ($\sim$19\%, $\sim$9\%, and $\sim$2\%, respectively). Note that NGC3079 has been counted twice due to the galaxy is confirmed as both disk and outflow maser types. Besides, there is also a group of putative masers, which including 6 putative disk masers, 3 putative jet/outflow masers and 3 putative star-forming masers ($\sim$14\%, $\sim$7\%, and $\sim$7\%, respectively). The rest of 19 masers remain unknown maser types. Our selected maser sample only represents $\sim$ 27\% of the MCP maser detected galaxies (161 targets). However, the disk maser fraction is consistent with previous studies (e.g \citealt{Kuo2018}). Besides, for five selected maser galaxies of NGC4922, NGC4253, NGC4051, NGC5256 and NGC5128, their spectra only range from 5.2 to 14.2 $\mu$m, while the rest reaches to 35 $\mu$m. Finally, all spectra of selected maser galaxies are listed in Appendix Figure~\ref{fig:spec_masers_all}. 

\begin{figure}
	\includegraphics[width=0.95\columnwidth]{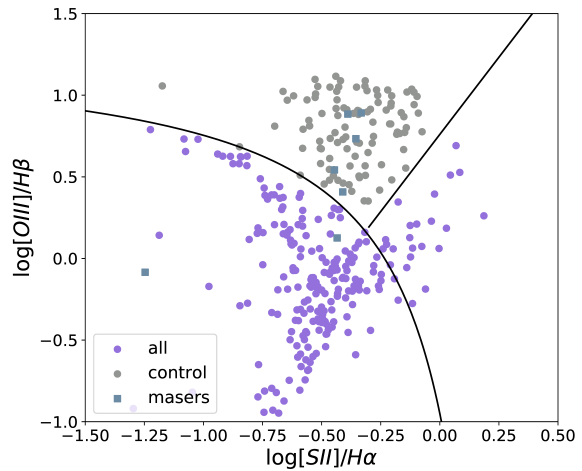}
    \caption{The final selected control sample of Seyfert 2 galaxies and its BPT diagram. The total number of selected Seyfert 2 galaxies is 78. NGC4194 is a star forming galaxy in our maser sample. }
    \label{fig:control_sample}
\end{figure}

\subsection{Control Sample of Seyfert 2 galaxies}
\label{sec: control}
The majority of maser galaxies have been found in the narrow emission line population, such as Seyfert 2 and a few LINERs. In our selected maser sample, there are 31 galaxies classified as Seyfert 2 (about three quarters; $\sim$72\%), which is the dominated population, after extensive searching from both NED and Simbad. Therefore, to improve our understanding of maser galaxies amid type 2 AGNs, in particular Seyfert 2 galaxies, we select a Seyfert 2 control sample with MIR emission line measurements.

The Seyfert 2 control sample is selected from the Sloan Digital Sky Survey (SDSS) Data Release 7 (DR7), where the emission line measurements of galaxy nuclei are adopted from the MPA-JHU DR7 catalogue \citep{Brinchmann2004}. We first cross matched the MIR spectral sample of \citet{Lambrides2019} and the MPA-JHU catalogue by a search radius of 3$\arcsec$, and obtained 338 non-maser galaxies with MIR spectra. This pre-selected sample includes both infrared and optical measurements. At the same time, we also removed the contamination from broad-line (type 1) AGNs based on the galaxy type classifications of MPA-JHU catalogue. 

The optical spectral classification of galaxies was conducted by using the BPT diagrams of  [S$\textrm{\sevensize{II}}$]/H$\alpha$ versus [O$\textrm{\sevensize{III}}$]/H$\beta$ and [N$\textrm{\sevensize{II}}$]/H$\alpha$ versus [O$\textrm{\sevensize{III}}$]/H$\beta$ (e.g. \citealt{Baldwin1981,Veilleux1987}). For the pre-selected galaxies, the SNRs of those five emission lines are all larger than 3. Galaxies above the `Kewley line' are classified as AGNs \citep{Kewley2001}. Moreover, the Seyfert 2 and LINERs are separated by the line adopted by \citet{Kewley2006}, where the Seyfert 2 population represents galaxies with higher excitation than LINERs. All selected control galaxies are located in the Seyfert 2 region based on the  [S$\textrm{\sevensize{II}}$]/H$\alpha$ versus [O$\textrm{\sevensize{III}}$]/H$\beta$ classification as shown in Figure~\ref{fig:control_sample}. We also compared the selected Seyfert 2 galaxies by using [N$\textrm{\sevensize{II}}$] or [S$\textrm{\sevensize{II}}$] to constraint the final Seyfert 2 control sample. The selection from [N$\textrm{\sevensize{II}}$]/H$\alpha$ contains 75 Seyfert 2 galaxies, while [S$\textrm{\sevensize{II}}$]/H$\alpha$ includes 78 galaxies. The galaxies selected by [N$\textrm{\sevensize{II}}$]/H$\alpha$ are all recovered by the [S$\textrm{\sevensize{II}}$]/H$\alpha$ criteria. Therefore, we adopt the [S$\textrm{\sevensize{II}}$]/H$\alpha$ as our selection criterion with 78 targets in the control sample.

\begin{figure}
	\includegraphics[width=\columnwidth]{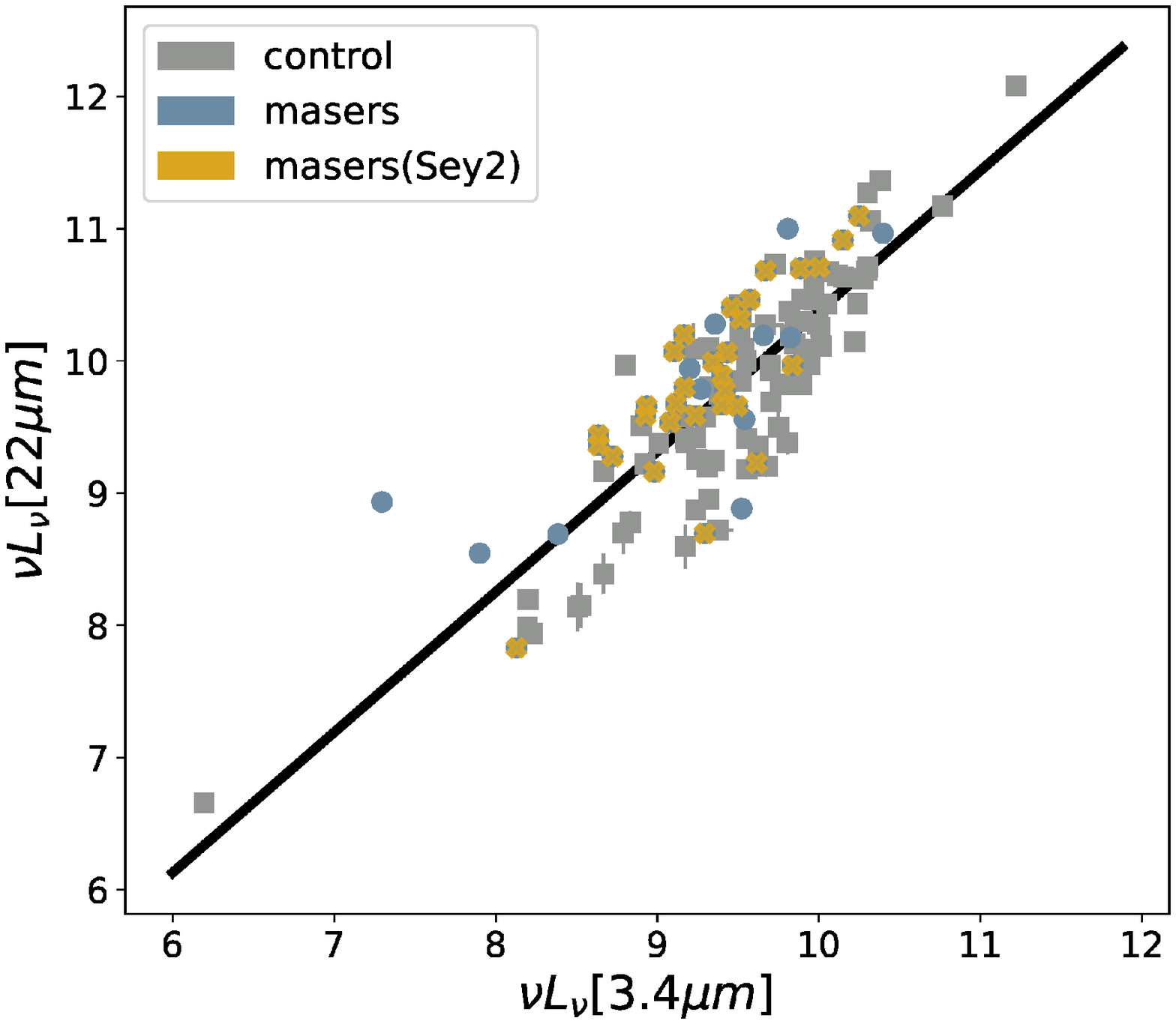} \\
	\includegraphics[width=\columnwidth]{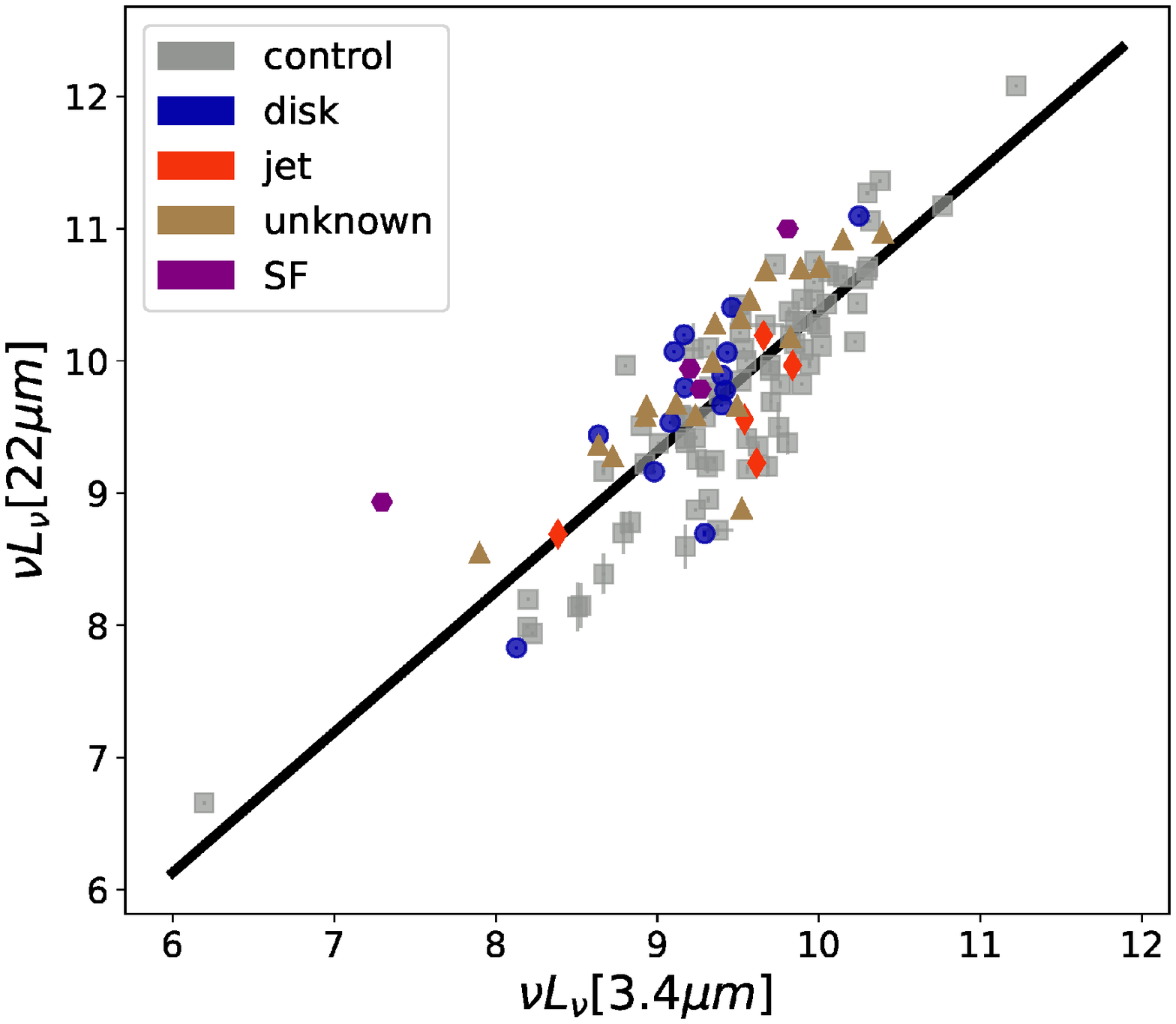}
    \caption{3.4 $\mu$m versus 22 $\mu$m luminosity for both masers galaxies and control sample. Generally, there is no significant difference between the two groups. Upper: A simple machine learning linear regression (solid line) has been applied to the combined sample (maser, Seyfert 2 maser sub-sample and control) to define a linear mean relation. Maser galaxies show mainly higher 22 $\mu$m luminosity compared to the control sample, while control sample galaxies could be separated into high- and low-22 $\mu$m populations.  Bottom: Maser galaxies are classified based on their maser types, which included both confirmed and putative disk, jet, star forming and unknown types and are labeled by different colours. No significant differences are found in different maser classes in this plot.  }
    \label{fig:samplecomparison}
\end{figure}

Moreover, in order to ensure our statistics results that were not introduced by the original selection effects in infrared luminosity, we compared the infrared photometric data of our maser galaxies and control sample, which was shown in upper panel of Figure~\ref{fig:samplecomparison} with no significant differences in mid-infrared luminosity between two samples. We also characterised the maser galaxies with different maser types and showed in the bottom panel of Figure \ref{fig:samplecomparison}, which also indicated no significant differences between maser galaxies and control sample.

\begin{table*}
	\centering
	\caption{Spitzer IRS module characteristics}
	\label{tab:IRS_table}
	\begin{tabular}{lcccr} 
		\hline\hline
Module & Wavelength Range & Spectral resolution & Slit Width & Slit Length \\
 & ($\mu$m) & (R = $\lambda/\Delta \lambda$) & ($\arcsec$) & ($\arcsec$) \\
\hline
SL & 5.2 - 14.5 & 60 < R < 128 & 3.6(3.7) & 57 \\
LL & 14 - 38 & 57 < R < 126 & 10.5(10.7) & 168 \\
SH & 9.9 - 19.6 & R $\sim$ 600 & 4.7 & 11.3 \\
LH & 18.7 - 37.2 & R $\sim$ 600 & 11.1 & 22.3 \\
		\hline\hline
	\end{tabular}
\end{table*}


\begin{table*}
	\centering
	\caption{Initial sample of 43 maser galaxies}
	\label{tab:masers_table}
	\begin{tabular}{lcccccccr} 
		\hline\hline
  AORkey$^{(a)}$ & Source Name & RA (J2000) & DEC (J2000) & ... & Err(PAH$_{11.3 \mu m}$)  & PAH$_{11.3 \mu m}$ (Upper Limit)$^{(b)}$ & $\tau_{9.7 \mu m}$ & maser class$^{(c)}$\\
        &              &  (deg)  & (deg) & ...& (10$^{39}$erg s$^{-1}$) & (10$^{41}$erg s$^{-1}$) &  &  \\
		\hline
  18968576  &  NGC291 & 13.374583 & -8.767778 & ...  & 0.1275076   &    ...    &  1.078882 &        /   \\
  25408000   &  Mrk1 & 19.030000 & 33.089444 & ...   & 0.1291172  & ... & 0.721732   &      D?   \\
   9073408   &      NGC520 & 21.146125 & 3.792417 & ...  & 0.02270371 & ... & 1.318774  &       SF? \\
  17798400   &  J0214-0016  & 33.524583  &  -0.276944 & ... & 0.2166990 &    ...  & 0.378285  &       / \\
   4850176   &   Mrk1066 & 44.994167 &  36.820556 & ...  & 0.1377199 &  ... & 0.816244    &     / \\
		\hline\hline
		\multicolumn{9}{l}{Note: A machine-readable full table is available on CDS. The first five targets are shown here in terms of its form and contents. } \\
		\multicolumn{9}{l}{$^{(a)}$ \emph{Spitzer} IRS identification number. } \\
	   \multicolumn{9}{l}{$^{(b)}$ Upper limit luminosity of PAH in 11.3 $\mu$m. If an emission line is $<$ 2$\sigma$, it will give an upper limit flux. More information can be found in \citet{Lambrides2019}.}  \\
	   \multicolumn{9}{l}{$^{(c)}$ Maser type classification adopted from \citet{Panessa2020} and the MCP website: D = disk; J = jet; M = multiple; SF = star-forming; / = unknown.  } \\
	   \multicolumn{9}{l}{When a `?' is added to the classification, it indicates that this is only putative/ not-confirmed. } \\

	\end{tabular}
\end{table*}

\begin{table}
	\centering
	\caption{Emission Line Detection Rate of Masers and Control Sample}
	\label{tab:detection_rate}
	
	\begin{tabular}{lcr} 
		\hline\hline
Emission Lines & Masers & Control \\
& (\%) & (\%) \\
\hline
PAH$_{6.2 \mu m}$ & 74 & 58 \\
PAH$_{7.7 \mu m}$ & 65 & 65 \\
PAH$_{11.3 \mu m}$ & 88 & 87 \\
H$_2$ S(1) & 74 & 45 \\
H$_2$ S(3) & 65 & 28 \\
$\tau_{9.7 \mu m}$ & 84 & 92 \\
		\hline\hline
	\end{tabular}
\end{table}

\section{Results}

Hereafter, we will investigate the MIR spectra, PAH emissions, silicate strength as well as the spectral indices of masers and control sample in this section. \\

\subsection{MIR Spectral Comparisons in Maser and Control Sample}

In order to directly show the difference of MIR spectral features between maser and control samples, we constructed the rest-frame median spectra of two samples. We first de-redshifted all our spectra to the rest-frame based on their redshifts or recessional velocities, and then resampled the spectra with the same wavelength intervals. After resampling, we normalized all selected spectra, and selected the median value in individual wavelength data point. Finally, we obtained our median spectra in maser and control samples (see Figure~\ref{fig:median_spectra}). In the upper panel of Figure~\ref{fig:median_spectra}, we presented the median spectra of maser and control Seyfert 2 galaxies from \emph{Spitzer}. As a comparison, we also plotted the starburst and Seyfert 1 galaxy templates adopted from \citet{Hernan-Caballero2011}. We found clearly that the MIR spectra of masers reside between typical starburst galaxies and Seyfert 1 galaxies, which is consistent with that dust grain can be destroyed by the hard radiation field. Moreover, the maser galaxies tend to show a much steeper spectral indices ($\alpha_{20-35 \mu m}$) than control sample. In the lower panel of Figure~\ref{fig:median_spectra}, we compared the spectral index of  maser galaxies to two control subsamples, which are separated by MIR luminosity (22 $\mu$m) based on Figure~\ref{fig:samplecomparison}. In this panel, we found that the maser galaxies showed a significant differences compared to low-22 $\mu$m luminosity (fainter) control sub-sample, but comparable with the high-22 $\mu$m luminosity (brighter) control sub-sample. Moreover, we also showed the median spectrum of maser Seyfert 2 sub-sample (Seyfert 2 galaxies in the maser sample), for which we still found it tended to have higher MIR luminosity. All these results indicate that maser galaxies ought to have high 22 $\mu$m luminosity.

\subsection{The Comparisons of PAH Emissions}

Polycyclic aromatic hydrocarbons (PAHs) are species of carbon-based molecules with the union of aromatic rings, which may be contributed the emission feature of 3.3, 6.2, 7.7, 8.6, 11.3, and 12.7 $\mu$m (e.g. \citealt{Li2020}). Most of previous studies suggested that PAH features were excited by circumnuclear kpc-scale star formation activity (e.g. \citealt{Smith2007}). However, many studies have shown controversial results (e.g. \citealt{Smith2007, Alonso-Herrero2014,Jensen2017}.

\begin{figure}
	\includegraphics[width=\columnwidth]{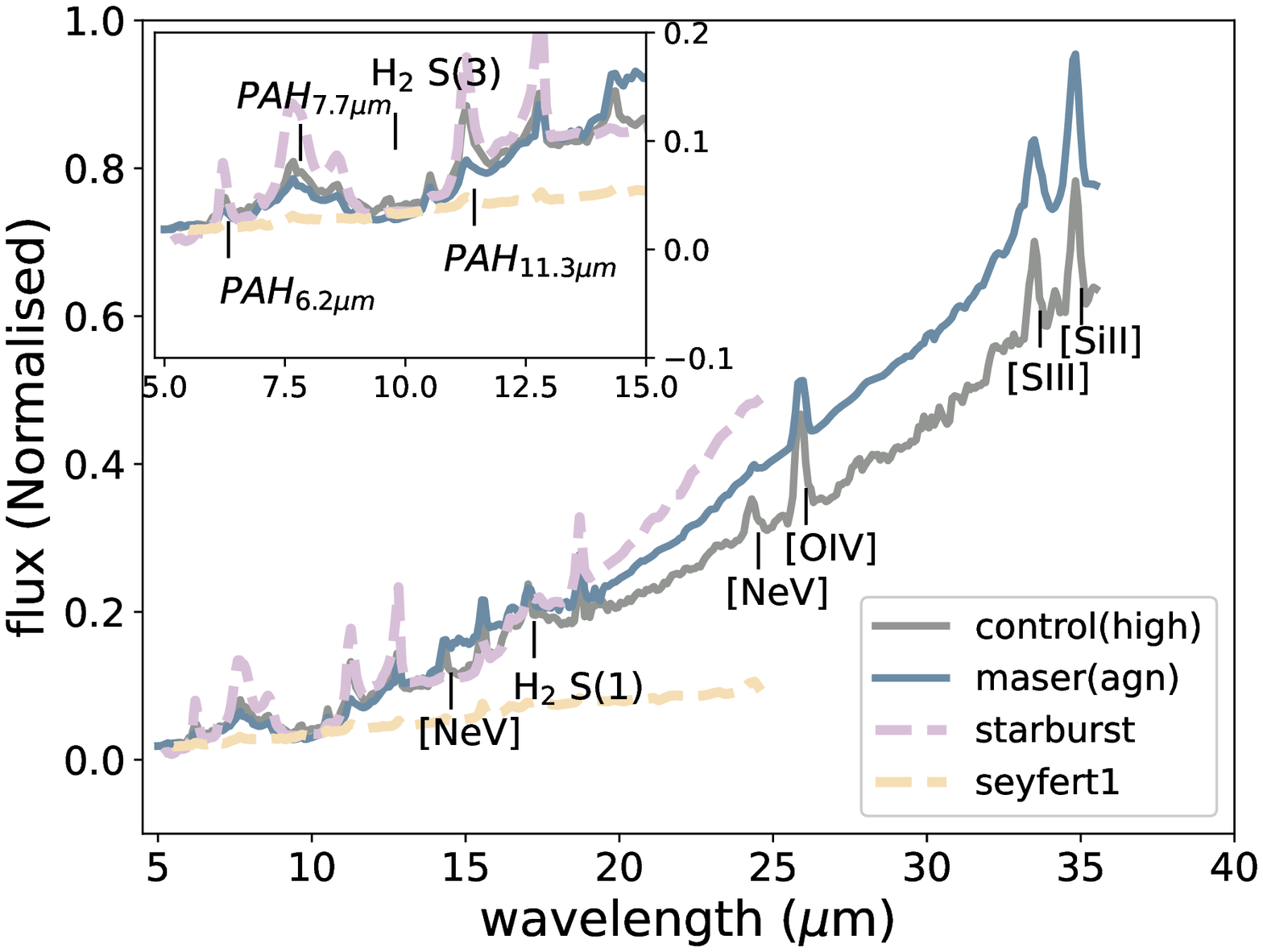} \\
	\includegraphics[width=\columnwidth]{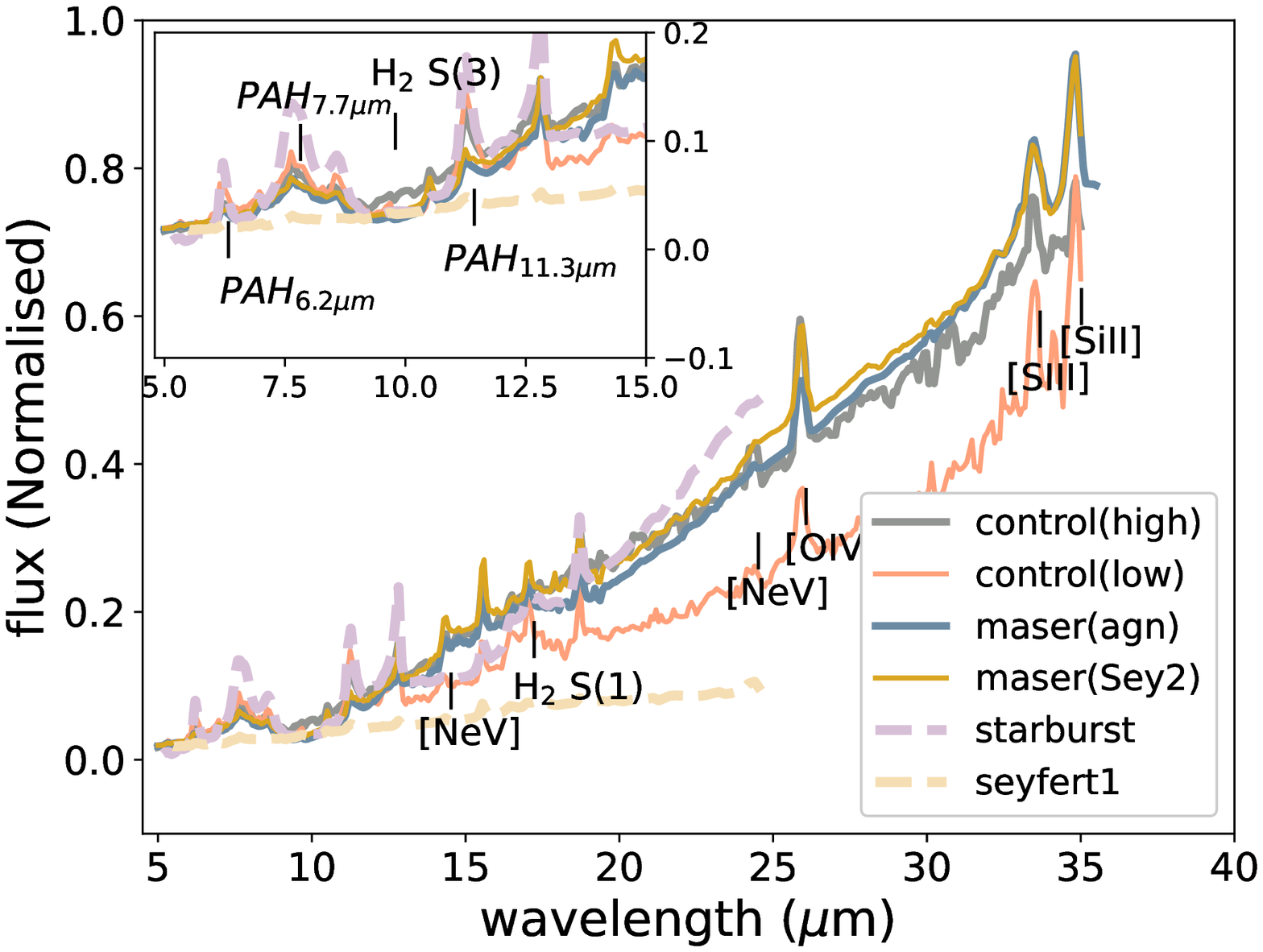}
    \caption{Upper: normalized median spectra of maser and control Seyfert 2 galaxies. The starburst and Seyfert 1 galaxy templates are adopted from \citet{Hernan-Caballero2011}.  Lower: normalized median spectra of maser and Seyfert 2 maser galaxies and high- and low-22$\mu$m luminosity control sample galaxies. Both insert panels show the zoom out region of 5 - 15 $\mu$m.}
    \label{fig:median_spectra}
\end{figure}

\begin{figure*}
	\includegraphics[width=5.5cm]{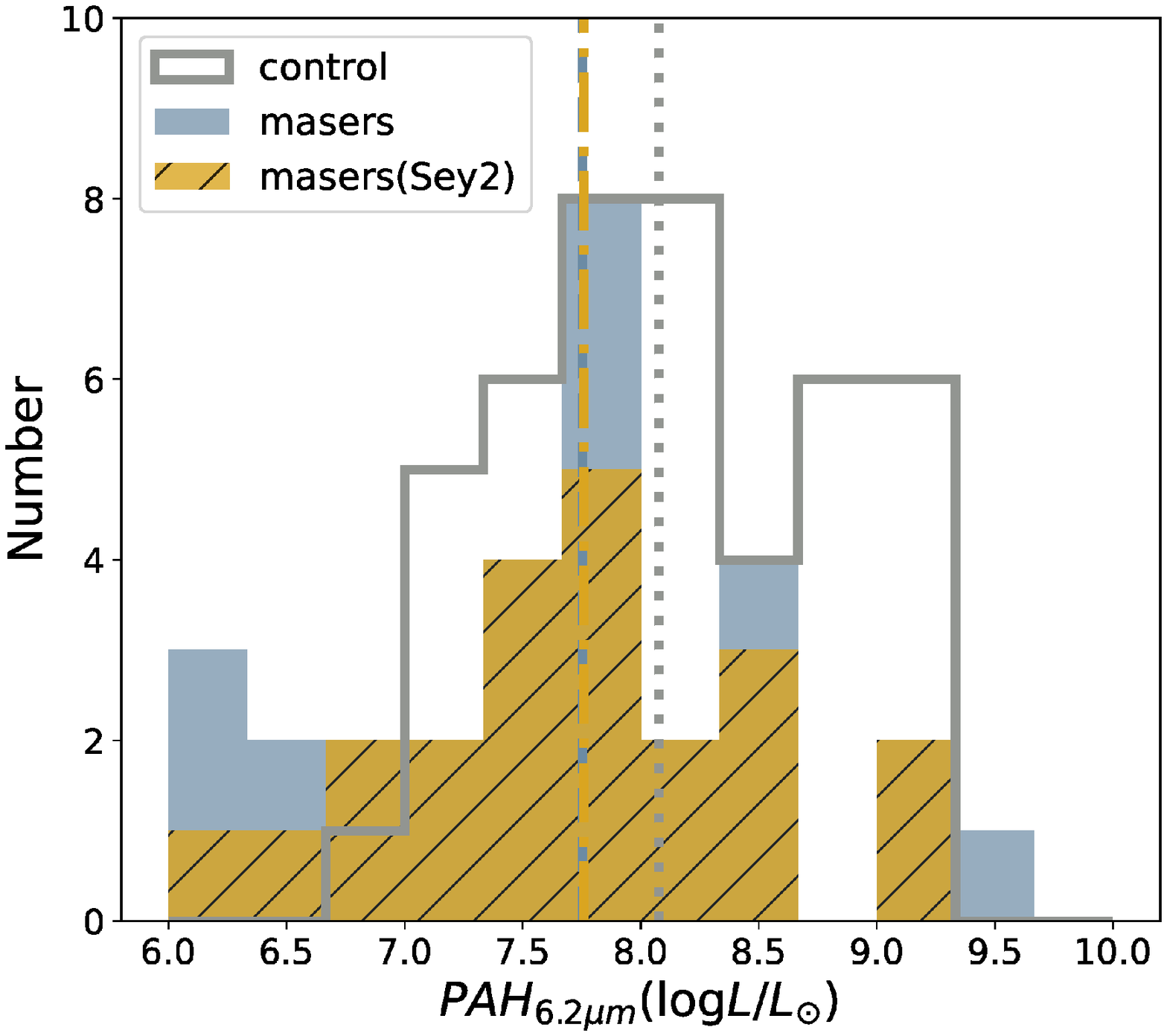}
	\includegraphics[width=5.5cm]{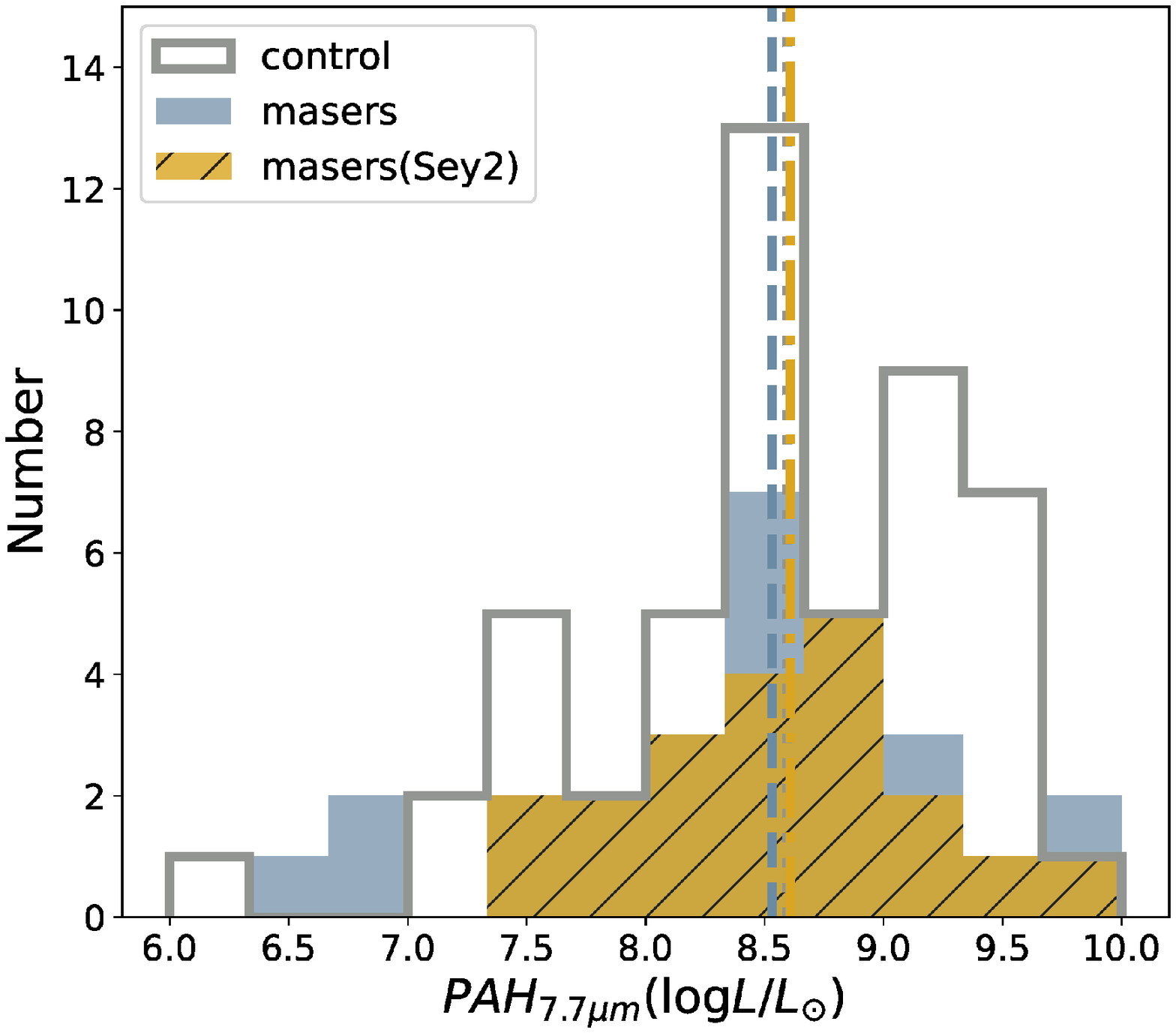}
	\includegraphics[width=5.5cm]{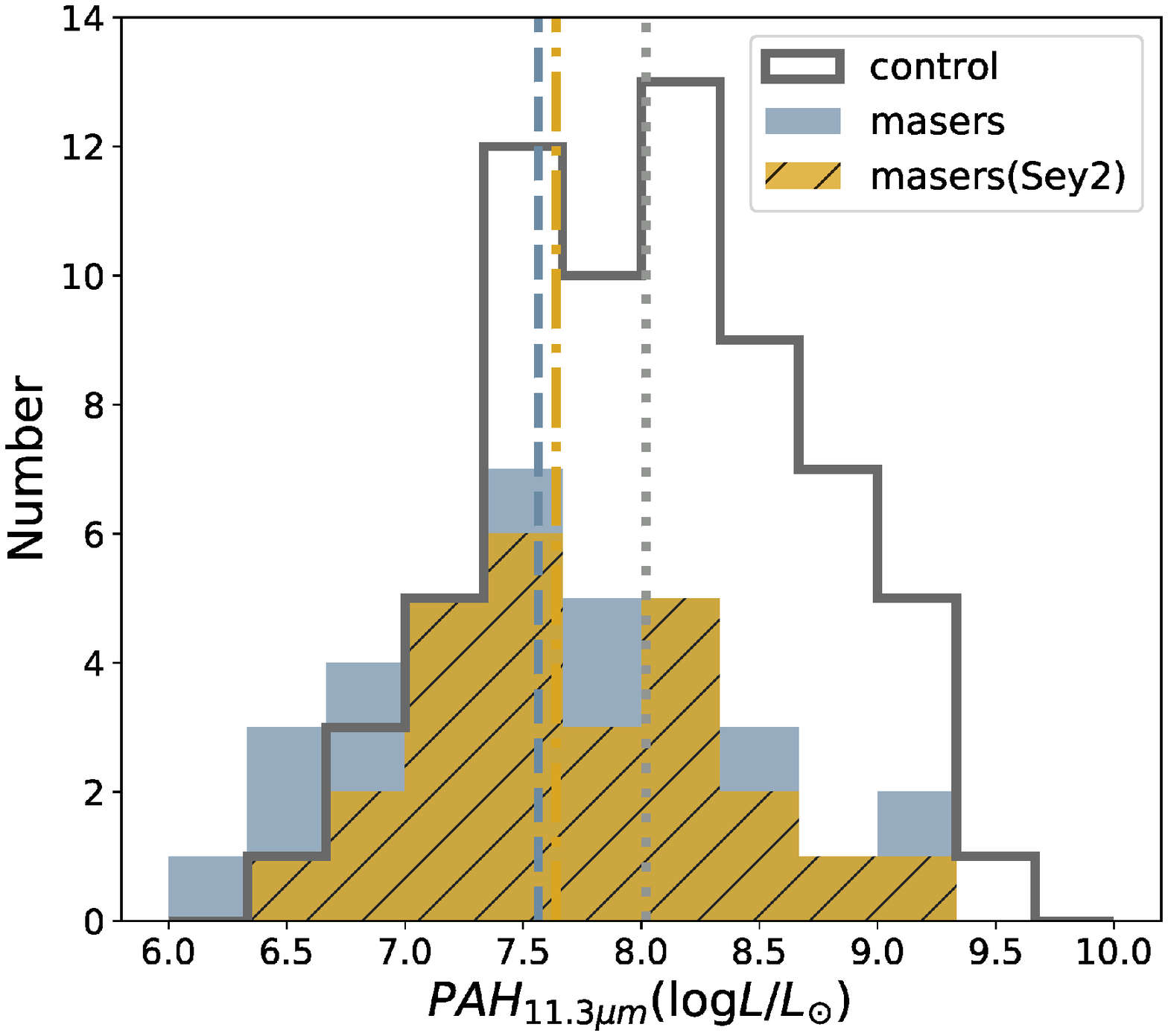}
    \caption{The histogram of PAH 6.2 (\textit{left}), 7.7 (\textit{middle}) and 11.3 $\mu$m (\textit{right}) luminosity between the whole maser sample (blue), Seyfert 2 maser sample (beige) and control sample galaxies (gray), respectively. The corresponding median values of maser sample and control sample are marked by dashed lines and dotted lines, respectively.}
    \label{fig:dist_pah_11_3}
\end{figure*}

In our study, we used the PAH luminosities which adopted from \citet{Lambrides2019} and then compared the PAH detection rates of 6.2 $\mu$m, 7.7 $\mu$m and 11.3 $\mu$m in masers and control sample galaxies as shown in Table~\ref{tab:detection_rate}. The detection rate of PAH 6.2 $\mu$m in masers is higher than in the control sample, while the detection rates of 7.7 $\mu$m and 11.3 $\mu$m are comparable between the two samples. However, the distribution of spectral feature strength at 6.2 $\mu$m and 7.7 $\mu$m (see Figure~\ref{fig:dist_pah_11_3}), which are formed by the PAH stretching mode, are also very similar and indistinguishable between the two samples (see also  Figure~\ref{fig:median_spectra}). Meanwhile, the histogram of 11.3 $\mu$m feature strength, which is excited by carbon-hydrogen modes, shows that the PAH 11.3 $\mu$m luminosity is lower in masers than in the control sample galaxies, as shown in Figure~\ref{fig:median_spectra}. Besides, in order to quantify the galaxy types effects, we also investigated the Seyfert 2 maser sub-sample, and found the PAH emissions at 6.2 $\mu$m and 7.7 $\mu$m are also no significant different from the control sample. However, the discrepancy in the PAH at 11.3 $\mu$m is even more signification between the Seyfert 2 maser sub-sample and control Seyfert 2 sample.   

\subsection{Mid-infrared Silicate Strength}

The silicate features at 9.7 $\mu$m and 18 $\mu$m are strong indicators of dust geometry as well as central engine activity (e.g. \citealt{Spoon2007,Hao2007}). For example, the presence of strong silicate absorption and absence of PAH emission indicates that the object is a heavily dust obscured AGN. Moreover, these silicate features are also particularly important for testing the unification model. 

\citet{Spoon2007} used the strength of the silicate 9.7 $\mu$m feature to quantitatively describe the galaxy types. Later studies also adopted the same measurement to study optical depth at the same wavelengths. The apparent depth of the silicate feature at 9.7 $\mu$m ($\tau_{9.7 \mu m}$) is inferred by adopting a local MIR continuum and evaluating the ratio of observed flux density ($f_{\textrm{obs}}$) with respect to continuum flux density ($f_{\textrm{cont}}$) at 9.7 $\mu$m. \citet{Lambrides2019} defined $\tau_{9.7 \mu m}$ as follows:
\begin{equation}
    \tau_{9.7 \mu m} = - \textrm{ln} \frac{f_{\textrm{obs}}[9.7 \mu m]}{f_{\textrm{cont}} [9.7 \mu m]}.
\end{equation}
Based on this definition, $\tau_{9.7 \mu m}$ and the silicate strength at 9.7$\mu$m ($S_{9.7 \mu m}$) can be converted to each other as shown below, so that we can compare our result to previous studies.

\begin{equation}
    \tau_{9.7 \mu m} = - S_{9.7 \mu m}
\end{equation}

\begin{figure}
	\includegraphics[width=0.9\columnwidth]{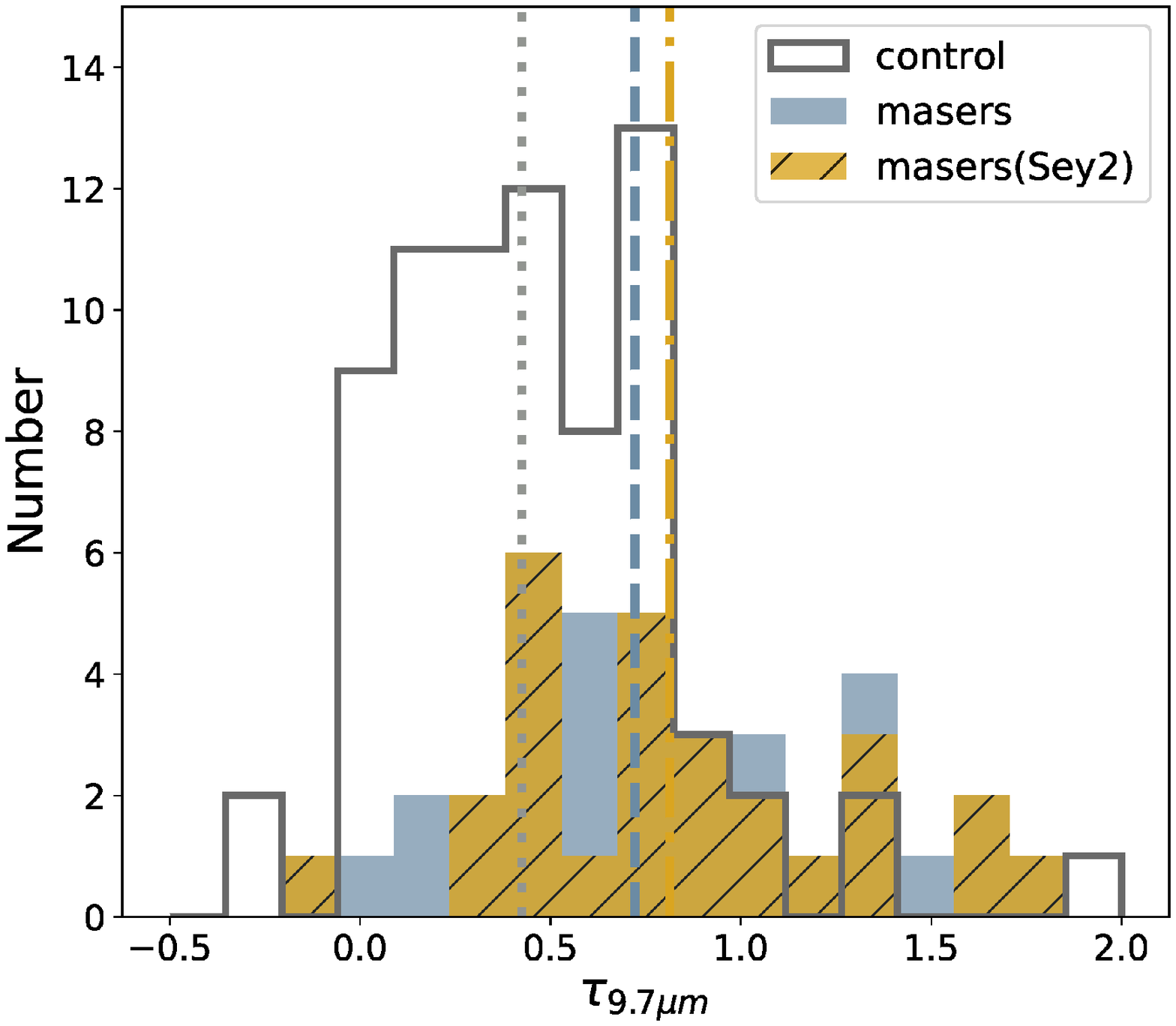}
    \caption{Histogram of $\tau_{9.7 \mu m}$ for maser sample (blue)), Seyfert 2 maser sample (beige) and control sample galaxies (gray). The corresponding median value of maser sample and control sample are marked by dashed line and dotted line, respectively.}
    \label{fig:dist_tau_10}
\end{figure}

In our sample, we found both maser and control sample galaxies show silicate absorption features at 9.7 $\mu$m. However, in the histogram of $\tau_{9.7 \mu m}$ (Figure~\ref{fig:dist_tau_10}), maser galaxies tend to have higher silicate absorption than the control sample. If taking only into account the Seyfert 2 maser sub-sample, the maser galaxies tend to show even higher silicate absorption with larger discrepancy between two samples. 

\subsection{Mid-infrared Spectral Index }

In order to compare generally the MIR spectra of maser galaxies and control sample, we calculated the spectral index, which is sensitive to the spectral continuum of galaxies. We defined the spectral index, which is unitless, as 

\begin{equation}
    \alpha_{12} = \frac{\log_{10} [f_1 (\lambda) / f_2 (\lambda)]}{\log_{10} [(\lambda_{1})/ (\lambda_{2})]}
\end{equation}

where $f(\lambda)$ and $\lambda$ is the flux density and corresponding wavelength of beginning and end of the continuum, respectively. As shown in the upper panel of Figure~\ref{fig:median_spectra}, the normalized median spectra (normalized to the maximum flux density) between maser galaxies and control sample have significantly different spectral indices at the wavelength range of 20 - 30 $\mu$m. When taking into account the Seyfert 2 maser sub-sample, the median spectral index reduce to be 1.68, only 4\% larger than the median value of control sample. This result is consistent with spectral index ranges of Seyfert 2 galaxies in \citet{Wu2009} and indicates that the Seyfert 2 maser sub-sample belongs to the MIR enhanced (above median) Seyfert 2 galaxies population. 

\begin{figure}
	\includegraphics[width=0.9\columnwidth]{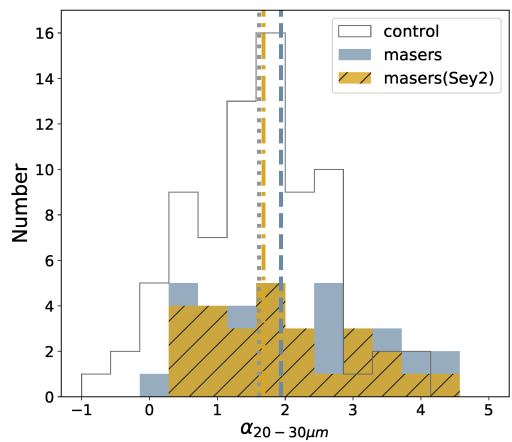}
    \caption{The distribution of spectral indices between maser and control sample galaxies. The median values of maser (dashed line) and control sample galaxies (dotted line) are 1.94 and 1.61, respectively. If only taking into account of the Seyfert 2 sub-sample, the median value reduce from 1.94 to 1.68, which is slightly larger than the control sample.}
    \label{fig:spectral_indices}
\end{figure}

\section{Discussions}

\subsection{The Detection of Extragalactic Water Masers}


Water masers are the most common found masers in the Universe. Disk masers are accurate indicators for black hole mass measurements as well as accurate distance indicators. Jet/outflow masers are useful to derive jet/outflows geometry as well as jet material speed (e.g. \citealt{Castangia2019,Surcis2020}). One of the major challenges for the utility of water masers is the small sample size currently available. Although $>$ 6000 galaxies in the nearby universe have been observed, the detection rate of masers remains low at only a few percent. Additionally, a question arises naturally: why do some galaxies host masers but the others do not? One way to address this question would be to compare the galaxies with detected masers and a control sample with similar properties. Since the presence of water molecules reveals a dusty environment \citep{Masini2016} and the excitation temperature of masers is required to be in the range of 300 - 1000 K, the MIR properties of masers and control galaxies may be able to reveal the intrinsic dust geometry as well as dust temperature in the maser torus.

Previous works have studied the multi-wavelength properties of masers. For example, in X-ray, AGNs in maser galaxies showed high obscuration or even Compton-thick disks (e.g. \citealt{Kuo2019, Masini2016}). Using optical spectra, \citet{Zhu2011} and \citet{Constantin2012} discovered a weak correlation between [O$\textrm{\sevensize{III}}$]$\lambda$5007 and maser luminosity. Moreover, \citet{Ramolla2011} found a significant difference between masers and non-masers for the X-ray/[O$\textrm{\sevensize{IV}}$]$_{25.9 \mu m}$ ratio, which was a hint of differences between masers and non-masers for the infrared spectra. \citet{Nenkova2002} created a simple clumpy torus model to describe the formation of the 10 $\mu$m absorption feature. In the model, they suggested that the 10 $\mu$m features in Seyfert 2 galaxies would present in absorption if the optically thick torus is in edge-on projection \citep{Roche1991}. In our sample, the silicate absorption feature of the entire control sample shows mild absorption, which is consistent with the unification model. Meanwhile, masers show slightly deeper absorption (see also Figure~\ref{fig:median_spectra}), indicating that the projection angles of masers may be smaller than for the control sample.

In addition, \citet{Kuo2018} pointed out that maser galaxies were associated with 22 $\mu$m luminosity enhancement. Therefore, we separated the control sample into high- and low-22$\mu$m luminosity sub-samples (Figure~\ref{fig:samplecomparison}). For the high-22 $\mu$m luminosity sample, the spectral index in 20 - 30$\mu$m becomes steeper and similar to maser galaxies (the lower panel of Figure~\ref{fig:median_spectra}), revealing that spectral indices of maser galaxies are not significantly different from control sample when 22 $\mu$m luminosity enhancement is taken into account. However, the 9.7 $\mu$m absorption feature of the control sample is shallower than that of masers. On the other hand, the 9.7 $\mu$m absorption features in low-22 $\mu$m luminosity sample has similar strength to that of maser galaxies, while the spectral indices show a remarkable difference (shallower) than maser galaxies. The steeper MIR spectral indices and deeper silicate absorption features, indicate that maser galaxies are more dust obscured than average Seyfert 2 galaxies. This is consistent with the results of \citet{Ramolla2011}, who suggested maser galaxies tend to be more dust obscured by studying the X-ray luminosity. 

Meanwhile, the high excitation emission lines in maser galaxies, such as [O$\textrm{\sevensize{IV}}$] and [Ne$\textrm{\sevensize{V}}$], are weaker than in the control sample. Previous studies found [Ne$\textrm{\sevensize{V}}$] and [O$\textrm{\sevensize{IV}}$] were mostly connected to AGNs. Therefore, the weaker high excitation emission lines may represent weaker central engines in maser galaxies. 

\subsection{PAH Emission in Water Masers}

The emissions at 6.2 $\mu$m and 7.7 $\mu$m are excited by radiative relaxation of carbon-carbon stretching modes, which represents the intensity of ionized PAH molecules. The strength of the 11.3 $\mu$m PAH feature, which is attributed to the carbon-hydrogen modes, could vary by one order of magnitude from completely ionized PAH to completely neutral PAH molecules. Previous studies found that the ratio between 6.2 $\mu$m and 7.7 $\mu$m PAH features showed insignificant variation when the ionization fraction changed (e.g. \citealt{Jensen2017}). The relative intensity between the two PAH bands is likely dependent on the distribution of grain size. The effect of AGNs on PAH intensities are still unclear.  Indeed, the AGNs show stronger radiation field and weaker PAH emissions than typical star formation galaxies. Some of studies suggest that the PAHs will be destroyed by the AGN radiation field (e.g. \citealt{Smith2007, Wu2010}) or dilute by the AGN continuum, while others find that the PAH intensity in AGN-dominated galaxies varies in a larger range but no destruction occurs (e.g. \citealt{Shipley2013, Stierwalt2014}). Recent studies found some hints that PAH might even survive in the nuclear environment (e.g. \citealt{Alonso-Herrero2014,Jensen2017}). 

We have found that masers galaxies can be distinguished from other Seyfert 2 galaxies by MIR PAH 11.3 $\mu$m as shown in Figure~\ref{fig:median_spectra} and Figure~\ref{fig:dist_pah_11_3}, where maser galaxies show significant low luminosity compared to control sample.  Therefore, further exploration of the 8-13 $\mu$m features could be an useful tool to enlarge the sample of maser candidates by using ground-based instruments. Moreover, the future space instrument will explore further information of maser galaxies.

\subsection{Maser Detection Rate}

Our analysis provides new avenues to improve maser detection rate using the MIR spectra, in particular some MIR indicators (e.g $\tau_{9.7 \mu m}$). Previous studies comparing non-masers and masers found that maser galaxies tended to have higher MIR luminosity (e.g. \citealt{Kuo2018}). However, there is no significant correlation between the maser emission and the AGN activity in multi-wavelength studies (e.g. \citealt{Zhu2011, Constantin2012, Castangia2013, Kuo2018}). Therefore, the question remains which parameters may be best suited to increase maser detection rates. 

According to \citet{Braatz2010}, the current maser detection rate is only $\sim$ 3\%. 
As mentioned in \citet{Kuo2018}, the masers tend to have an enhancement at 22 $\mu$m, which is higher than 10$^9$ L$_{\odot}$. Therefore, \citet{Kuo2018} suggested that if only 22 $\mu$m enhanced galaxies are selected, then the detection rate of maser galaxies would increase to about $6\sim18$\%. Moreover, the maser detection rate could increase significantly in a criterion exploiting MIR and X-ray sample \citep{Castangia2016,Castangia2019}. \citet{Panessa2020} found the maser detection rate tend to be much higher in type 2 and Compton-thick AGNs (Further discussion in \S 5.4). In this studies, we focus on improving maser detection rate using MIR spectra. 

We quote useful numbers for calculating detection rates of masers in Tables \ref{tab:selection} and \ref{tab:selection_combination}. Raw percentages of masers in our total MIR detected sample (masers and control) are 36\% overall and 50\% for high-22 $\mu$m objects. However, these numbers may not reflect actual detection rates as they are dominated by the \textit{Spitzer} spectra. This includes both random effects (due to scientific interests of \textit{Spitzer} proposals) and a preference for objects with high MIR fluxes that could be effectively observed by \textit{Spitzer}. 

\begin{table*}
	\centering
	\caption{Numbers of objects useful for calculating detection rates}
	\label{tab:selection}
	
	\begin{tabular}{lcccccc} 
		\hline\hline
Sample & N$_{\mathrm{tot}}$ &  N$_{\mathrm{withIR}}$ &  N$_{\mathrm{high22}}$  &  N$_{\mathrm{low22}}$ & $N_{\tau_{9.7 \mu m}}$ > 0.8 & $N_{\mathrm{PAH}_{11.3 \mu m}}$ < 7.5\\
& (\%) & (\%) \\
\hline
MCP database & 4836 & 58 & - & - & - & -\\
Masers & 161 & 43 & 32 & 11 & 17 & 18\\
Control & - & 76 & 35 & 43 & 9 & 15\\
		\hline\hline
	\end{tabular}
\end{table*}

\begin{table}
	\centering
	\caption{Numbers of objects useful for calculating detection rates (combined indicators)}
	\label{tab:selection_combination}
\begin{tabular}{lcc} 
		\hline\hline
Sample (high22) & $N_{\tau_{9.7 \mu m} > 0.8}$ & $N_{\mathrm{PAH}_{11.3 \mu m}}$ < 7.5\\
\hline
Masers & 10 & 12\\
Control & 3 & 2\\

		\hline\hline
	\end{tabular}
\end{table}

We attempt to correct for this selection effect by assuming that our selected maser and control sample are representative in the MCP project. Then we can scale the number of objects in the maser sample and the control sample to the number of masers and non-maser galaxies in the MCP by using factors of 3.7 and 75, respectively. If we select the high-22$\mu$m sub-sample, the maser detection rate increases to $\sim$ 5\%. This is a lower limit to the actual detection rate, as our MIR-selected sample is already 22 $\mu$m bright by definition. This number is thus consistent with the result of \citet{Kuo2018}. 

More indicators which are sensitive to the presence of masers in galaxies may also increase the detection rate. In particular, those mid-infrared indicators between 8 - 13 $\mu$m (such as $\tau_{9.7 \mu m}$) could be observed through ground-based telescope. These indicators may become important to increase the detection rate of masers and thus the overall maser sample in an effective way. We define a second mid-infrared criterion as $\tau_{9.7 \mu m} > 0.8$. In the high-22$\mu$m sub-sample, $\sim$ 28.5\% of our selected maser galaxies show $\tau_{9.7 \mu m} > 0.8$. In contrast, only $\sim$ 9\% of control sample galaxies show a similar $\tau_{9.7 \mu m}$ enhancement. In other words, if we combine the 22 $\mu$m enhancement and the 9.7 $\mu$m opacity, the detection rate may increase to $\sim$16\% in the global maser detected sample (currently: 180/$\sim$6000), which is a factor of 5 higher than the current detection rate. 

 We note that, alternatively, the PAH emission feature at 11.3 $\mu$m seems to be another sensitive indicator. As can be seen from Tables~\ref{tab:selection} and \ref{tab:selection_combination}, the PAH 11.3 $\mu$m increases the detection rate just as much as $\tau_{9.7 \mu m}$, which generally indicates that PAH 11.3 $\mu$m emission tends to be lower in maser galaxies. Indeed, the detection rate combining high 22 $\mu$m emission and low PAH 11.3 $\mu$m would nominally be 29\% for our samples. However, sample statistics in the control sample start to be really low (just 2 objects). Also, the PAH emission may be contaminated by star formation activity. Therefore, we would recommend aiming for the $\tau_{9.7 \mu m}$ enhancement in the high-22 $\mu$m luminosity sample to significantly increase the maser detection rate.

\subsection{The Comparisons Between X-ray and MIR selection }
One major question which is still unclear is the excitation conditions of water maser emissions in galaxies. According to theoretical model, the central AGNs could heat the circumnuclear gas to be suitable for maser emissions. Under this assumption, X-ray luminosity may connect to maser luminosity. However, previous studies found controversial results. For example, \citet{Kondratko2006b} found such correlation with large scatter, while \citet{Castangia2013} found the correlation was absent when only taking into account the disk megamaser sources. Furthermore, \citet{Castangia2016, Castangia2019} found the maser detection rate increased significantly (up to 50\%) in a criterion exploiting MIR and X-ray sample. \citet{Panessa2020} studied an unbiased AGN sample, for which they found the detection rate significantly increased when considering type 2 and Compton-thick AGNs. The detection rate reached to 56$\pm$ 18\% in Compton-thick AGN sample. Besides, they also revisited the correlation between maser and X-ray luminosity, and confirmed the correlation did no exist.

The infrared luminosity usually comes from dust grain heated by UV photon. Based on the AGN unified model, the dusty torus is orientation-dependent for the central engine obscuration \citep{Antonucci1993}. \citet{Tristram2014} found the temperature of infrared emissions are consistent with dust at $T \sim$ 300 K, which is also comparable with water maser excitation temperature (e.g. \citealt{Lo2005}). Previous studies suggested that the dust and the maser disk are co-spatial in some maser host galaxies (i.e. NGC1068 and Circinus; \citealt{Raban2009, Tristram2014}). In our study, we cross-matched our maser sample with \citet{Panessa2020}, where 19 common galaxies were found. The median column density of these common galaxies is $\log (N_{\textrm{H}}) = 23.82$, which is in agree with the results that maser galaxies tend to be in highly absorbed population (e.g. \citealt{Castangia2019, Panessa2020}). The rest maser galaxies in our sample may be also in highly obscured population due to the large X-ray luminosity uncertainty. Besides, the detection rate in our infrared selected sample increased to be $\sim$ 40 \% if taking into account the criteria of $\tau_{9.7 \mu m} > 0.8$ or $\mathrm{PAH}_{11.3 \mu m}$ < 7.5 in the current selected sample (43 maser galaxies in total, see Table~\ref{tab:selection}), which significantly improve the maser detection rate.

\subsection{Sample Selection Bias and Error Budgets}

Generally, selection biases in the context of maser galaxies (or even in a broader sense, for AGNs) mainly depend on the requirement to measure luminosities or spectral features of the central AGNs. However, the selection is not determined by a single parameter. For example, a valid measurement of luminosity requires host galaxies to be brighter than the sensitivity of the observation (i.e., $\rm SNR>3$). On the other hand, absorption or emission feature measurements require the host galaxies to either have a significant activity or to be sufficiently massive. In that sense, multiple factors may affect the detection of maser galaxies.

For our sample, the major limitation is the completeness of infrared spectra from \textit{Spitzer}, for which only about 27\% of the known extragalactic maser sources reported by the MCP are covered by \textit{Spitzer} observation. According to Figure~\ref{fig:samplecomparison} and \ref{fig:median_spectra}, the maser and control Seyfert 2 galaxies do not show a remarkable difference between each other at high 22 $\mu$m luminosities. Therefore, the luminosity selection bias can be neglected. Moreover, as shown in Figure~\ref{fig:samplecomparison}, the different maser types show more or less comparable characteristics that there is no significant difference among them. This indicates that the infrared luminosity may not easily distinguished the maser types.

The potential mid-infrared indicators, such as PAH$_{11.3\mu m}$ and $\tau_{9.7\mu m}$ are shown significant differences between masers and control sample and are not affected by galaxy spectral types, even if only taking into account the Seyfert 2 maser sub-sample. For instance, the median values of $\tau_{9.7\mu m}$ in maser sample and control sample are 0.72 and 0.42, respectively. Meanwhile, the median values of $\tau_{9.7\mu m}$ in Seyfert 2 maser sub-sample is 0.81, which is even larger than the median value of the entire maser sample.

Meanwhile, the mid-infrared spectral index is affected by galaxy spectral types and might not be a sensitive maser indicator. If we take the whole maser sample (regardless the galaxy spectral type) into account, the maser spectral indices are largely different from the control sample (see Figure~\ref{fig:spectral_indices}). However, as expected, the Seyfert 2 masers have no significant difference from the control Seyfert 2 sample, as the median spectral indices vary from 1.94 to 1.68 in whole maser galaxies and Seyfert 2 maser sub-sample, respectively. This variation is reasonable since spectral index is a continuum slope indicator. 

\section{Summary}

Water megamasers at 22 GHz are a great tool for investigating the gas dynamics of the innermost region of AGN, since they are direct dynamical indicators. However, a major challenge for water megamasers is to increase the sample size by improving the very low detection rate. Since the excitation temperature of water megamasers is about 300 K $<$ T $<$ 1000 K, which is in the MIR wavelength range, we take a close look at the MIR spectral properties in maser and control Seyfert 2 galaxies, and find the following results:

\begin{itemize}
    \item The PAH 11.3 $\mu$m emission and silicate absorption $\tau_{9.7 \mu m}$ may be two potential indicators of water maser emission in galaxies, as there are differences between maser and control samples as well as maser Seyfert 2 subsample in these two indices. Also, these same spectral features can be observed in future space- or ground-based instruments. We estimate that the detection rate could be increased from 3\% currently to 16\%, which is a factor 5 increasing, when using the $\tau_{9.7 \mu m}$ feature in the 22 $\mu$m enhancement sample. It could be further increased to 28.5\% when using the PAH 11.3 $\mu$m emission in the 22 $\mu$m enhancement sample, which would be a factor 10 increasing in detection rate. 
    \item Besides, our result of MIR spectral slope $\alpha_{20-30 \mu m}$ suggested that masers and control Seyfert 2 galaxies may have different slope, and the dust temperatures of masers galaxies may be cooler than in control Seyfert 2 galaxies. While the maser Seyfert 2 subsample are not so sensitive to this indicator but still belongs to MIR enhanced Seyfert 2 population.

\end{itemize}

Although the masers tend to be a MIR enhanced population in this study, the faint end of the galaxy luminosity function at 22 $\mu$m (or 3.4$\mu$m) may still contain masers. Future instruments such as JWST-MIRI or other mid-infrared instruments may be useful for enlarging the sample size of maser galaxies and answer the question.

\section*{Acknowledgements}

We would like to thank the anonymous referee for many constructive comments and suggestions. M. Y. gratefully acknowledge the support from  the European Research Council (ERC) under the European Union’s Horizon 2020 research and innovation programme (grant agreement number 772086). 

This work is based in part on observations made with the \emph{Spitzer} Space Telescope, which is operated by the Jet Propulsion Laboratory, California Institute of Technology under a contract with NASA. This publication makes use of data product from the Wide-field Infrared Survey Explorer (WISE), which is a joint project of the University of California, Los Angeles, and the Jet Propulsion Laboratory/California Institute of Technology. It is funded by the National Aeronautics and Space Administration. This publication makes use of data from \emph{AllWISE} programme, which is funded by the NASA Science Mission Directorate Astrophysics Division and overseen by the NASA Explorer Office at Goddard Space Flight Center. Prof. Edward Wright of the University of California Los Angeles if the \emph{AllWISE} Principal Investigator. \emph{AllWISE} is managed by the Jet Propulsion Laboratory, California Institute of Technology. Data processing, archiving and distribution for \emph{AllWISE} is carried out by the Infrared Processing and Analysis Center, California Institute of Technology.

This publication makes use of data product from SDSS. Funding for the SDSS and SDSS-II has been provided by the Alfred P. Sloan Foundation, the Participating Institutions, the National Science Foundation, the U.S. Department of Energy, the National Aeronautics and Space Administration, the Japanese Monbukagakusho, the Max Planck Society, and the Higher Education Funding Council for England. The SDSS Web Site is http://www.sdss.org/.

The SDSS is managed by the Astrophysical Research Consortium for the Participating Institutions. The Participating Institutions are the American Museum of Natural History, Astrophysical Institute Potsdam, University of Basel, University of Cambridge, Case Western Reserve University, University of Chicago, Drexel University, Fermilab, the Institute for Advanced Study, the Japan Participation Group, Johns Hopkins University, the Joint Institute for Nuclear Astrophysics, the Kavli Institute for Particle Astrophysics and Cosmology, the Korean Scientist Group, the Chinese Academy of Sciences (LAMOST), Los Alamos National Laboratory, the Max-Planck-Institute for Astronomy (MPIA), the Max-Planck-Institute for Astrophysics (MPA), New Mexico State University, Ohio State University, University of Pittsburgh, University of Portsmouth, Princeton University, the United States Naval Observatory, and the University of Washington.

\section*{Data Availability}
The data underlying this article are available in the article and in its online supplementary material.




\bibliographystyle{mnras}
\bibliography{example} 

\begin{thebibliography}{}
\makeatletter
\relax
\def\mn@urlcharsother{\let\do\@makeother \do\$\do\&\do\#\do\^\do\_\do\%\do\~}
\def\mn@doi{\begingroup\mn@urlcharsother \@ifnextchar [ {\mn@doi@}
  {\mn@doi@[]}}
\def\mn@doi@[#1]#2{\def\@tempa{#1}\ifx\@tempa\@empty \href
  {http://dx.doi.org/#2} {doi:#2}\else \href {http://dx.doi.org/#2} {#1}\fi
  \endgroup}
\def\mn@eprint#1#2{\mn@eprint@#1:#2::\@nil}
\def\mn@eprint@arXiv#1{\href {http://arxiv.org/abs/#1} {{\tt arXiv:#1}}}
\def\mn@eprint@dblp#1{\href {http://dblp.uni-trier.de/rec/bibtex/#1.xml}
  {dblp:#1}}
\def\mn@eprint@#1:#2:#3:#4\@nil{\def\@tempa {#1}\def\@tempb {#2}\def\@tempc
  {#3}\ifx \@tempc \@empty \let \@tempc \@tempb \let \@tempb \@tempa \fi \ifx
  \@tempb \@empty \def\@tempb {arXiv}\fi \@ifundefined
  {mn@eprint@\@tempb}{\@tempb:\@tempc}{\expandafter \expandafter \csname
  mn@eprint@\@tempb\endcsname \expandafter{\@tempc}}}

\bibitem[\protect\citeauthoryear{{Alonso-Herrero} et~al.,}{{Alonso-Herrero}
  et~al.}{2014}]{Alonso-Herrero2014}
{Alonso-Herrero} A.,  et~al., 2014, \mn@doi [\mnras] {10.1093/mnras/stu1293},
  \href {https://ui.adsabs.harvard.edu/abs/2014MNRAS.443.2766A} {443, 2766}

\bibitem[\protect\citeauthoryear{{Antonucci}}{{Antonucci}}{1993}]{Antonucci1993}
{Antonucci} R.,  1993, \mn@doi [\araa] {10.1146/annurev.aa.31.090193.002353},
  \href {https://ui.adsabs.harvard.edu/abs/1993ARA&A..31..473A} {31, 473}

\bibitem[\protect\citeauthoryear{{Assef} et~al.,}{{Assef}
  et~al.}{2013}]{Assef2013}
{Assef} R.~J.,  et~al., 2013, \mn@doi [\apj] {10.1088/0004-637X/772/1/26},
  \href {https://ui.adsabs.harvard.edu/abs/2013ApJ...772...26A} {772, 26}

\bibitem[\protect\citeauthoryear{{Baldwin}, {Phillips}  \&
  {Terlevich}}{{Baldwin} et~al.}{1981}]{Baldwin1981}
{Baldwin} J.~A.,  {Phillips} M.~M.,   {Terlevich} R.,  1981, \mn@doi [\pasp]
  {10.1086/130766}, \href
  {https://ui.adsabs.harvard.edu/abs/1981PASP...93....5B} {93, 5}

\bibitem[\protect\citeauthoryear{{Braatz}, {Wilson}  \& {Henkel}}{{Braatz}
  et~al.}{1997}]{Braatz1997}
{Braatz} J.~A.,  {Wilson} A.~S.,   {Henkel} C.,  1997, \mn@doi [\apjs]
  {10.1086/312999}, \href
  {https://ui.adsabs.harvard.edu/abs/1997ApJS..110..321B} {110, 321}

\bibitem[\protect\citeauthoryear{{Braatz}, {Reid}, {Humphreys}, {Henkel},
  {Condon}  \& {Lo}}{{Braatz} et~al.}{2010}]{Braatz2010}
{Braatz} J.~A.,  {Reid} M.~J.,  {Humphreys} E.~M.~L.,  {Henkel} C.,  {Condon}
  J.~J.,   {Lo} K.~Y.,  2010, \mn@doi [\apj] {10.1088/0004-637X/718/2/657},
  \href {https://ui.adsabs.harvard.edu/abs/2010ApJ...718..657B} {718, 657}

\bibitem[\protect\citeauthoryear{{Braatz} et~al.,}{{Braatz}
  et~al.}{2018}]{Braatz2018}
{Braatz} J.,  et~al., 2018, in {Tarchi} A.,  {Reid} M.~J.,   {Castangia} P.,
  eds,  IAU Symposium Vol. 336, Astrophysical Masers: Unlocking the Mysteries
  of the Universe. pp 86--91, \mn@doi{10.1017/S1743921317010249}

\bibitem[\protect\citeauthoryear{{Brinchmann}, {Charlot}, {White}, {Tremonti},
  {Kauffmann}, {Heckman}  \& {Brinkmann}}{{Brinchmann}
  et~al.}{2004}]{Brinchmann2004}
{Brinchmann} J.,  {Charlot} S.,  {White} S.~D.~M.,  {Tremonti} C.,  {Kauffmann}
  G.,  {Heckman} T.,   {Brinkmann} J.,  2004, \mn@doi [\mnras]
  {10.1111/j.1365-2966.2004.07881.x}, \href
  {https://ui.adsabs.harvard.edu/abs/2004MNRAS.351.1151B} {351, 1151}

\bibitem[\protect\citeauthoryear{{Castangia}, {Panessa}, {Henkel}, {Kadler}  \&
  {Tarchi}}{{Castangia} et~al.}{2013}]{Castangia2013}
{Castangia} P.,  {Panessa} F.,  {Henkel} C.,  {Kadler} M.,   {Tarchi} A.,
  2013, \mn@doi [\mnras] {10.1093/mnras/stt1824}, \href
  {https://ui.adsabs.harvard.edu/abs/2013MNRAS.436.3388C} {436, 3388}

\bibitem[\protect\citeauthoryear{{Castangia}, {Tarchi}, {Caccianiga},
  {Severgnini}  \& {Della Ceca}}{{Castangia} et~al.}{2016}]{Castangia2016}
{Castangia} P.,  {Tarchi} A.,  {Caccianiga} A.,  {Severgnini} P.,   {Della
  Ceca} R.,  2016, \mn@doi [\aap] {10.1051/0004-6361/201527177}, \href
  {https://ui.adsabs.harvard.edu/abs/2016A&A...586A..89C} {586, A89}

\bibitem[\protect\citeauthoryear{{Castangia}, {Surcis}, {Tarchi}, {Caccianiga},
  {Severgnini}  \& {Della Ceca}}{{Castangia} et~al.}{2019}]{Castangia2019}
{Castangia} P.,  {Surcis} G.,  {Tarchi} A.,  {Caccianiga} A.,  {Severgnini} P.,
    {Della Ceca} R.,  2019, \mn@doi [\aap] {10.1051/0004-6361/201935421}, \href
  {https://ui.adsabs.harvard.edu/abs/2019A&A...629A..25C} {629, A25}

\bibitem[\protect\citeauthoryear{{Constantin}}{{Constantin}}{2012}]{Constantin2012}
{Constantin} A.,  2012, in Journal of Physics Conference Series. p. 012047
  (\mn@eprint {arXiv} {1201.3925}), \mn@doi{10.1088/1742-6596/372/1/012047}

\bibitem[\protect\citeauthoryear{{Gao} et~al.,}{{Gao} et~al.}{2016}]{Gao2016}
{Gao} F.,  et~al., 2016, \mn@doi [\apj] {10.3847/0004-637X/817/2/128}, \href
  {https://ui.adsabs.harvard.edu/abs/2016ApJ...817..128G} {817, 128}

\bibitem[\protect\citeauthoryear{{Greenhill}, {Jiang}, {Moran}, {Reid}, {Lo}
  \& {Claussen}}{{Greenhill} et~al.}{1995}]{Greenhill1995}
{Greenhill} L.~J.,  {Jiang} D.~R.,  {Moran} J.~M.,  {Reid} M.~J.,  {Lo} K.~Y.,
   {Claussen} M.~J.,  1995, \mn@doi [\apj] {10.1086/175301}, \href
  {https://ui.adsabs.harvard.edu/abs/1995ApJ...440..619G} {440, 619}

\bibitem[\protect\citeauthoryear{{Greenhill} et~al.,}{{Greenhill}
  et~al.}{2003}]{Greenhill2003}
{Greenhill} L.~J.,  et~al., 2003, \mn@doi [\apj] {10.1086/374862}, \href
  {https://ui.adsabs.harvard.edu/abs/2003ApJ...590..162G} {590, 162}

\bibitem[\protect\citeauthoryear{{Greenhill}, {Tilak}  \&
  {Madejski}}{{Greenhill} et~al.}{2008}]{Greenhill2008}
{Greenhill} L.~J.,  {Tilak} A.,   {Madejski} G.,  2008, \mn@doi [\apjl]
  {10.1086/592782}, \href
  {https://ui.adsabs.harvard.edu/abs/2008ApJ...686L..13G} {686, L13}

\bibitem[\protect\citeauthoryear{{Hagiwara}, {Diamond}, {Miyoshi}, {Rovilos}
  \& {Baan}}{{Hagiwara} et~al.}{2003}]{Hagiwara2003}
{Hagiwara} Y.,  {Diamond} P.~J.,  {Miyoshi} M.,  {Rovilos} E.,   {Baan} W.,
  2003, \mn@doi [\mnras] {10.1046/j.1365-8711.2003.07005.x}, \href
  {https://ui.adsabs.harvard.edu/abs/2003MNRAS.344L..53H} {344, L53}

\bibitem[\protect\citeauthoryear{{Hao}, {Weedman}, {Spoon}, {Marshall},
  {Levenson}, {Elitzur}  \& {Houck}}{{Hao} et~al.}{2007}]{Hao2007}
{Hao} L.,  {Weedman} D.~W.,  {Spoon} H.~W.~W.,  {Marshall} J.~A.,  {Levenson}
  N.~A.,  {Elitzur} M.,   {Houck} J.~R.,  2007, \mn@doi [\apjl]
  {10.1086/511973}, \href
  {https://ui.adsabs.harvard.edu/abs/2007ApJ...655L..77H} {655, L77}

\bibitem[\protect\citeauthoryear{{Henkel}, {Greene}  \& {Kamali}}{{Henkel}
  et~al.}{2018}]{Henkel2018}
{Henkel} C.,  {Greene} J.~E.,   {Kamali} F.,  2018, in {Tarchi} A.,  {Reid}
  M.~J.,   {Castangia} P.,  eds,  IAU Symposium Vol. 336, Astrophysical Masers:
  Unlocking the Mysteries of the Universe. pp 69--79 (\mn@eprint {arXiv}
  {1802.04727}), \mn@doi{10.1017/S1743921318000753}

\bibitem[\protect\citeauthoryear{{Hern{\'a}n-Caballero} \&
  {Hatziminaoglou}}{{Hern{\'a}n-Caballero} \&
  {Hatziminaoglou}}{2011}]{Hernan-Caballero2011}
{Hern{\'a}n-Caballero} A.,  {Hatziminaoglou} E.,  2011, \mn@doi [\mnras]
  {10.1111/j.1365-2966.2011.18413.x}, \href
  {https://ui.adsabs.harvard.edu/abs/2011MNRAS.414..500H} {414, 500}

\bibitem[\protect\citeauthoryear{{Humphreys}, {Reid}, {Moran}, {Greenhill}  \&
  {Argon}}{{Humphreys} et~al.}{2013}]{Humphreys2013}
{Humphreys} E.~M.~L.,  {Reid} M.~J.,  {Moran} J.~M.,  {Greenhill} L.~J.,
  {Argon} A.~L.,  2013, \mn@doi [\apj] {10.1088/0004-637X/775/1/13}, \href
  {https://ui.adsabs.harvard.edu/abs/2013ApJ...775...13H} {775, 13}

\bibitem[\protect\citeauthoryear{{Jensen} et~al.,}{{Jensen}
  et~al.}{2017}]{Jensen2017}
{Jensen} J.~J.,  et~al., 2017, \mn@doi [\mnras] {10.1093/mnras/stx1447}, \href
  {https://ui.adsabs.harvard.edu/abs/2017MNRAS.470.3071J} {470, 3071}

\bibitem[\protect\citeauthoryear{{Kewley}, {Dopita}, {Sutherland}, {Heisler}
  \& {Trevena}}{{Kewley} et~al.}{2001}]{Kewley2001}
{Kewley} L.~J.,  {Dopita} M.~A.,  {Sutherland} R.~S.,  {Heisler} C.~A.,
  {Trevena} J.,  2001, \mn@doi [\apj] {10.1086/321545}, \href
  {https://ui.adsabs.harvard.edu/abs/2001ApJ...556..121K} {556, 121}

\bibitem[\protect\citeauthoryear{{Kewley}, {Groves}, {Kauffmann}  \&
  {Heckman}}{{Kewley} et~al.}{2006}]{Kewley2006}
{Kewley} L.~J.,  {Groves} B.,  {Kauffmann} G.,   {Heckman} T.,  2006, \mn@doi
  [\mnras] {10.1111/j.1365-2966.2006.10859.x}, \href
  {https://ui.adsabs.harvard.edu/abs/2006MNRAS.372..961K} {372, 961}

\bibitem[\protect\citeauthoryear{{Kondratko} et~al.,}{{Kondratko}
  et~al.}{2006a}]{Kondratko2006a}
{Kondratko} P.~T.,  et~al., 2006a, \mn@doi [\apj] {10.1086/498641}, \href
  {https://ui.adsabs.harvard.edu/abs/2006ApJ...638..100K} {638, 100}

\bibitem[\protect\citeauthoryear{{Kondratko}, {Greenhill}  \&
  {Moran}}{{Kondratko} et~al.}{2006b}]{Kondratko2006b}
{Kondratko} P.~T.,  {Greenhill} L.~J.,   {Moran} J.~M.,  2006b, \mn@doi [\apj]
  {10.1086/507885}, \href
  {https://ui.adsabs.harvard.edu/abs/2006ApJ...652..136K} {652, 136}

\bibitem[\protect\citeauthoryear{{Kuo} et~al.,}{{Kuo} et~al.}{2011}]{Kuo2011}
{Kuo} C.~Y.,  et~al., 2011, \mn@doi [\apj] {10.1088/0004-637X/727/1/20}, \href
  {https://ui.adsabs.harvard.edu/abs/2011ApJ...727...20K} {727, 20}

\bibitem[\protect\citeauthoryear{{Kuo}, {Braatz}, {Reid}, {Lo}, {Condon},
  {Impellizzeri}  \& {Henkel}}{{Kuo} et~al.}{2013}]{Kuo2013}
{Kuo} C.~Y.,  {Braatz} J.~A.,  {Reid} M.~J.,  {Lo} K.~Y.,  {Condon} J.~J.,
  {Impellizzeri} C.~M.~V.,   {Henkel} C.,  2013, \mn@doi [\apj]
  {10.1088/0004-637X/767/2/155}, \href
  {https://ui.adsabs.harvard.edu/abs/2013ApJ...767..155K} {767, 155}

\bibitem[\protect\citeauthoryear{{Kuo} et~al.,}{{Kuo} et~al.}{2018}]{Kuo2018}
{Kuo} C.~Y.,  et~al., 2018, \mn@doi [\apj] {10.3847/1538-4357/aac498}, \href
  {https://ui.adsabs.harvard.edu/abs/2018ApJ...860..169K} {860, 169}

\bibitem[\protect\citeauthoryear{{Kuo} et~al.,}{{Kuo} et~al.}{2019}]{Kuo2019}
{Kuo} C.~Y.,  et~al., 2019, arXiv e-prints, \href
  {https://ui.adsabs.harvard.edu/abs/2019arXiv191110721K} {p. arXiv:1911.10721}

\bibitem[\protect\citeauthoryear{{Kuo} et~al.,}{{Kuo} et~al.}{2020}]{Kuo2020}
{Kuo} C.~Y.,  et~al., 2020, \mn@doi [\apj] {10.3847/1538-4357/ab781d}, \href
  {https://ui.adsabs.harvard.edu/abs/2020ApJ...892...18K} {892, 18}

\bibitem[\protect\citeauthoryear{{Kylafis} \& {Norman}}{{Kylafis} \&
  {Norman}}{1991}]{Kylafis1991}
{Kylafis} N.~D.,  {Norman} C.~A.,  1991, \mn@doi [\apj] {10.1086/170071}, \href
  {https://ui.adsabs.harvard.edu/abs/1991ApJ...373..525K} {373, 525}

\bibitem[\protect\citeauthoryear{{Lacy} et~al.,}{{Lacy}
  et~al.}{2004}]{Lacy2004}
{Lacy} M.,  et~al., 2004, \mn@doi [\apjs] {10.1086/422816}, \href
  {https://ui.adsabs.harvard.edu/abs/2004ApJS..154..166L} {154, 166}

\bibitem[\protect\citeauthoryear{{Lambrides}, {Petric}, {Tchernyshyov},
  {Zakamska}  \& {Watts}}{{Lambrides} et~al.}{2019}]{Lambrides2019}
{Lambrides} E.~L.,  {Petric} A.~O.,  {Tchernyshyov} K.,  {Zakamska} N.~L.,
  {Watts} D.~J.,  2019, \mn@doi [\mnras] {10.1093/mnras/stz1316}, \href
  {https://ui.adsabs.harvard.edu/abs/2019MNRAS.487.1823L} {487, 1823}

\bibitem[\protect\citeauthoryear{{Li}}{{Li}}{2020}]{Li2020}
{Li} A.,  2020, \mn@doi [Nature Astronomy] {10.1038/s41550-020-1051-1}, \href
  {https://ui.adsabs.harvard.edu/abs/2020NatAs...4..339L} {4, 339}

\bibitem[\protect\citeauthoryear{{Lo}}{{Lo}}{2005}]{Lo2005}
{Lo} K.~Y.,  2005, \mn@doi [\araa] {10.1146/annurev.astro.41.011802.094927},
  \href {https://ui.adsabs.harvard.edu/abs/2005ARA&A..43..625L} {43, 625}

\bibitem[\protect\citeauthoryear{{Maloney}}{{Maloney}}{2002}]{Maloney2002}
{Maloney} P.~R.,  2002, \mn@doi [\pasa] {10.1071/AS01100}, \href
  {https://ui.adsabs.harvard.edu/abs/2002PASA...19..401M} {19, 401}

\bibitem[\protect\citeauthoryear{{Masini} et~al.,}{{Masini}
  et~al.}{2016}]{Masini2016}
{Masini} A.,  et~al., 2016, \mn@doi [\aap] {10.1051/0004-6361/201527689}, \href
  {https://ui.adsabs.harvard.edu/abs/2016A&A...589A..59M} {589, A59}

\bibitem[\protect\citeauthoryear{{Miyoshi}, {Moran}, {Herrnstein}, {Greenhill},
  {Nakai}, {Diamond}  \& {Inoue}}{{Miyoshi} et~al.}{1995}]{Miyoshi1995}
{Miyoshi} M.,  {Moran} J.,  {Herrnstein} J.,  {Greenhill} L.,  {Nakai} N.,
  {Diamond} P.,   {Inoue} M.,  1995, \mn@doi [\nat] {10.1038/373127a0}, \href
  {https://ui.adsabs.harvard.edu/abs/1995Natur.373..127M} {373, 127}

\bibitem[\protect\citeauthoryear{{Nenkova}, {Ivezi{\'c}}  \&
  {Elitzur}}{{Nenkova} et~al.}{2002}]{Nenkova2002}
{Nenkova} M.,  {Ivezi{\'c}} {\v{Z}}.,   {Elitzur} M.,  2002, \mn@doi [\apjl]
  {10.1086/340857}, \href
  {https://ui.adsabs.harvard.edu/abs/2002ApJ...570L...9N} {570, L9}

\bibitem[\protect\citeauthoryear{{Panessa}, {Castangia}, {Malizia}, {Bassani},
  {Tarchi}, {Bazzano}  \& {Ubertini}}{{Panessa} et~al.}{2020}]{Panessa2020}
{Panessa} F.,  {Castangia} P.,  {Malizia} A.,  {Bassani} L.,  {Tarchi} A.,
  {Bazzano} A.,   {Ubertini} P.,  2020, \mn@doi [\aap]
  {10.1051/0004-6361/201937407}, \href
  {https://ui.adsabs.harvard.edu/abs/2020A&A...641A.162P} {641, A162}

\bibitem[\protect\citeauthoryear{{Peck}, {Henkel}, {Ulvestad}, {Brunthaler},
  {Falcke}, {Elitzur}, {Menten}  \& {Gallimore}}{{Peck}
  et~al.}{2003}]{Peck2003}
{Peck} A.~B.,  {Henkel} C.,  {Ulvestad} J.~S.,  {Brunthaler} A.,  {Falcke} H.,
  {Elitzur} M.,  {Menten} K.~M.,   {Gallimore} J.~F.,  2003, \mn@doi [\apj]
  {10.1086/374924}, \href
  {https://ui.adsabs.harvard.edu/abs/2003ApJ...590..149P} {590, 149}

\bibitem[\protect\citeauthoryear{{Pesce}, {Braatz}, {Condon}, {Gao}, {Henkel},
  {Litzinger}, {Lo}  \& {Reid}}{{Pesce} et~al.}{2015}]{Pesce2015}
{Pesce} D.~W.,  {Braatz} J.~A.,  {Condon} J.~J.,  {Gao} F.,  {Henkel} C.,
  {Litzinger} E.,  {Lo} K.~Y.,   {Reid} M.~J.,  2015, \mn@doi [\apj]
  {10.1088/0004-637X/810/1/65}, \href
  {https://ui.adsabs.harvard.edu/abs/2015ApJ...810...65P} {810, 65}

\bibitem[\protect\citeauthoryear{{Pesce} et~al.,}{{Pesce}
  et~al.}{2020}]{Pesce2020}
{Pesce} D.~W.,  et~al., 2020, \mn@doi [\apjl] {10.3847/2041-8213/ab75f0}, \href
  {https://ui.adsabs.harvard.edu/abs/2020ApJ...891L...1P} {891, L1}

\bibitem[\protect\citeauthoryear{{Raban}, {Jaffe}, {R{\"o}ttgering},
  {Meisenheimer}  \& {Tristram}}{{Raban} et~al.}{2009}]{Raban2009}
{Raban} D.,  {Jaffe} W.,  {R{\"o}ttgering} H.,  {Meisenheimer} K.,   {Tristram}
  K. R.~W.,  2009, \mn@doi [\mnras] {10.1111/j.1365-2966.2009.14439.x}, \href
  {https://ui.adsabs.harvard.edu/abs/2009MNRAS.394.1325R} {394, 1325}

\bibitem[\protect\citeauthoryear{{Ramolla}, {Haas}, {Bennert}  \&
  {Chini}}{{Ramolla} et~al.}{2011}]{Ramolla2011}
{Ramolla} M.,  {Haas} M.,  {Bennert} V.~N.,   {Chini} R.,  2011, \mn@doi [\aap]
  {10.1051/0004-6361/201015247}, \href
  {https://ui.adsabs.harvard.edu/abs/2011A&A...530A.147R} {530, A147}

\bibitem[\protect\citeauthoryear{{Reid}, {Braatz}, {Condon}, {Lo}, {Kuo},
  {Impellizzeri}  \& {Henkel}}{{Reid} et~al.}{2013}]{Reid2013}
{Reid} M.~J.,  {Braatz} J.~A.,  {Condon} J.~J.,  {Lo} K.~Y.,  {Kuo} C.~Y.,
  {Impellizzeri} C.~M.~V.,   {Henkel} C.,  2013, \mn@doi [\apj]
  {10.1088/0004-637X/767/2/154}, \href
  {https://ui.adsabs.harvard.edu/abs/2013ApJ...767..154R} {767, 154}

\bibitem[\protect\citeauthoryear{{Roche}, {Aitken}, {Smith}  \& {Ward}}{{Roche}
  et~al.}{1991}]{Roche1991}
{Roche} P.~F.,  {Aitken} D.~K.,  {Smith} C.~H.,   {Ward} M.~J.,  1991, \mn@doi
  [\mnras] {10.1093/mnras/248.4.606}, \href
  {https://ui.adsabs.harvard.edu/abs/1991MNRAS.248..606R} {248, 606}

\bibitem[\protect\citeauthoryear{{Severgnini}, {Caccianiga}  \& {Della
  Ceca}}{{Severgnini} et~al.}{2012}]{Severgnini2012}
{Severgnini} P.,  {Caccianiga} A.,   {Della Ceca} R.,  2012, \mn@doi [\aap]
  {10.1051/0004-6361/201118417}, \href
  {https://ui.adsabs.harvard.edu/abs/2012A&A...542A..46S} {542, A46}

\bibitem[\protect\citeauthoryear{{Shipley}, {Papovich}, {Rieke}, {Dey},
  {Jannuzi}, {Moustakas}  \& {Weiner}}{{Shipley} et~al.}{2013}]{Shipley2013}
{Shipley} H.~V.,  {Papovich} C.,  {Rieke} G.~H.,  {Dey} A.,  {Jannuzi} B.~T.,
  {Moustakas} J.,   {Weiner} B.,  2013, \mn@doi [\apj]
  {10.1088/0004-637X/769/1/75}, \href
  {https://ui.adsabs.harvard.edu/abs/2013ApJ...769...75S} {769, 75}

\bibitem[\protect\citeauthoryear{{Smith} et~al.,}{{Smith}
  et~al.}{2007}]{Smith2007}
{Smith} J.~D.~T.,  et~al., 2007, \mn@doi [\apj] {10.1086/510549}, \href
  {https://ui.adsabs.harvard.edu/abs/2007ApJ...656..770S} {656, 770}

\bibitem[\protect\citeauthoryear{{Spoon}, {Marshall}, {Houck}, {Elitzur},
  {Hao}, {Armus}, {Brandl}  \& {Charmandaris}}{{Spoon}
  et~al.}{2007}]{Spoon2007}
{Spoon} H.~W.~W.,  {Marshall} J.~A.,  {Houck} J.~R.,  {Elitzur} M.,  {Hao} L.,
  {Armus} L.,  {Brandl} B.~R.,   {Charmandaris} V.,  2007, \mn@doi [\apjl]
  {10.1086/511268}, \href
  {https://ui.adsabs.harvard.edu/abs/2007ApJ...654L..49S} {654, L49}

\bibitem[\protect\citeauthoryear{{Stern} et~al.,}{{Stern}
  et~al.}{2012}]{Stern2012}
{Stern} D.,  et~al., 2012, \mn@doi [\apj] {10.1088/0004-637X/753/1/30}, \href
  {https://ui.adsabs.harvard.edu/abs/2012ApJ...753...30S} {753, 30}

\bibitem[\protect\citeauthoryear{{Stierwalt} et~al.,}{{Stierwalt}
  et~al.}{2014}]{Stierwalt2014}
{Stierwalt} S.,  et~al., 2014, \mn@doi [\apj] {10.1088/0004-637X/790/2/124},
  \href {https://ui.adsabs.harvard.edu/abs/2014ApJ...790..124S} {790, 124}

\bibitem[\protect\citeauthoryear{{Surcis}, {Tarchi}  \& {Castangia}}{{Surcis}
  et~al.}{2020}]{Surcis2020}
{Surcis} G.,  {Tarchi} A.,   {Castangia} P.,  2020, \mn@doi [\aap]
  {10.1051/0004-6361/201937380}, \href
  {https://ui.adsabs.harvard.edu/abs/2020A&A...637A..57S} {637, A57}

\bibitem[\protect\citeauthoryear{{Tarchi}}{{Tarchi}}{2012}]{Tarchi2012}
{Tarchi} A.,  2012, in {Booth} R.~S.,  {Vlemmings} W. H.~T.,   {Humphreys} E.
  M.~L.,  eds,  IAU Symposium Vol. 287, Cosmic Masers - from OH to H0. pp
  323--332 (\mn@eprint {arXiv} {1205.3623}), \mn@doi{10.1017/S1743921312007259}

\bibitem[\protect\citeauthoryear{{Tarchi}, {Castangia}, {Columbano}, {Panessa}
  \& {Braatz}}{{Tarchi} et~al.}{2011}]{Tarchi2011}
{Tarchi} A.,  {Castangia} P.,  {Columbano} A.,  {Panessa} F.,   {Braatz} J.~A.,
   2011, \mn@doi [\aap] {10.1051/0004-6361/201117213}, \href
  {https://ui.adsabs.harvard.edu/abs/2011A&A...532A.125T} {532, A125}

\bibitem[\protect\citeauthoryear{{Tristram}, {Burtscher}, {Jaffe},
  {Meisenheimer}, {H{\"o}nig}, {Kishimoto}, {Schartmann}  \&
  {Weigelt}}{{Tristram} et~al.}{2014}]{Tristram2014}
{Tristram} K. R.~W.,  {Burtscher} L.,  {Jaffe} W.,  {Meisenheimer} K.,
  {H{\"o}nig} S.~F.,  {Kishimoto} M.,  {Schartmann} M.,   {Weigelt} G.,  2014,
  \mn@doi [\aap] {10.1051/0004-6361/201322698}, \href
  {https://ui.adsabs.harvard.edu/abs/2014A&A...563A..82T} {563, A82}

\bibitem[\protect\citeauthoryear{{Veilleux} \& {Osterbrock}}{{Veilleux} \&
  {Osterbrock}}{1987}]{Veilleux1987}
{Veilleux} S.,  {Osterbrock} D.~E.,  1987, \mn@doi [\apjs] {10.1086/191166},
  \href {https://ui.adsabs.harvard.edu/abs/1987ApJS...63..295V} {63, 295}

\bibitem[\protect\citeauthoryear{{Wright} et~al.,}{{Wright}
  et~al.}{2010}]{Wright2010}
{Wright} E.~L.,  et~al., 2010, \mn@doi [\aj] {10.1088/0004-6256/140/6/1868},
  \href {https://ui.adsabs.harvard.edu/abs/2010AJ....140.1868W} {140, 1868}

\bibitem[\protect\citeauthoryear{{Wu}, {Charmandaris}, {Huang}, {Spinoglio}  \&
  {Tommasin}}{{Wu} et~al.}{2009}]{Wu2009}
{Wu} Y.,  {Charmandaris} V.,  {Huang} J.,  {Spinoglio} L.,   {Tommasin} S.,
  2009, \mn@doi [\apj] {10.1088/0004-637X/701/1/658}, \href
  {https://ui.adsabs.harvard.edu/abs/2009ApJ...701..658W} {701, 658}

\bibitem[\protect\citeauthoryear{{Wu} et~al.,}{{Wu} et~al.}{2010}]{Wu2010}
{Wu} Y.,  et~al., 2010, \mn@doi [\apj] {10.1088/0004-637X/723/1/895}, \href
  {https://ui.adsabs.harvard.edu/abs/2010ApJ...723..895W} {723, 895}

\bibitem[\protect\citeauthoryear{{Yan} et~al.,}{{Yan} et~al.}{2013}]{Yan2013}
{Yan} L.,  et~al., 2013, \mn@doi [\aj] {10.1088/0004-6256/145/3/55}, \href
  {https://ui.adsabs.harvard.edu/abs/2013AJ....145...55Y} {145, 55}

\bibitem[\protect\citeauthoryear{{Yates}, {Field}  \& {Gray}}{{Yates}
  et~al.}{1997}]{Yates1997}
{Yates} J.~A.,  {Field} D.,   {Gray} M.~D.,  1997, \mn@doi [\mnras]
  {10.1093/mnras/285.2.303}, \href
  {https://ui.adsabs.harvard.edu/abs/1997MNRAS.285..303Y} {285, 303}

\bibitem[\protect\citeauthoryear{{Zhang}, {Henkel}, {Guo}, {Wang}  \&
  {Fan}}{{Zhang} et~al.}{2010}]{Zhang2010}
{Zhang} J.~S.,  {Henkel} C.,  {Guo} Q.,  {Wang} H.~G.,   {Fan} J.~H.,  2010,
  \mn@doi [\apj] {10.1088/0004-637X/708/2/1528}, \href
  {https://ui.adsabs.harvard.edu/abs/2010ApJ...708.1528Z} {708, 1528}

\bibitem[\protect\citeauthoryear{{Zhang}, {Henkel}, {Guo}  \& {Wang}}{{Zhang}
  et~al.}{2012}]{Zhang2012}
{Zhang} J.~S.,  {Henkel} C.,  {Guo} Q.,   {Wang} J.,  2012, \mn@doi [\aap]
  {10.1051/0004-6361/201117946}, \href
  {https://ui.adsabs.harvard.edu/abs/2012A&A...538A.152Z} {538, A152}

\bibitem[\protect\citeauthoryear{{Zhang}, {Liu}  \& {Henkel}}{{Zhang}
  et~al.}{2018}]{Zhang2018}
{Zhang} J.~S.,  {Liu} Z.~W.,   {Henkel} C.,  2018, in {Tarchi} A.,  {Reid}
  M.~J.,   {Castangia} P.,  eds,  IAU Symposium Vol. 336, Astrophysical Masers:
  Unlocking the Mysteries of the Universe. pp 92--95 (\mn@eprint {arXiv}
  {1712.06083}), \mn@doi{10.1017/S1743921317009814}

\bibitem[\protect\citeauthoryear{{Zhu}, {Zaw}, {Blanton}  \& {Greenhill}}{{Zhu}
  et~al.}{2011}]{Zhu2011}
{Zhu} G.,  {Zaw} I.,  {Blanton} M.~R.,   {Greenhill} L.~J.,  2011, \mn@doi
  [\apj] {10.1088/0004-637X/742/2/73}, \href
  {https://ui.adsabs.harvard.edu/abs/2011ApJ...742...73Z} {742, 73}

\makeatother
\end{thebibliography}




\appendix

\section{Maser Catalogue}

\textbf{Table~\ref{tab:masers_table}} The full machine-readable table is attached in the electron form. 

\section{Mid-infrared Spectra of All Selected Masers}
\begin{figure*}
	\includegraphics[width=17.6cm]{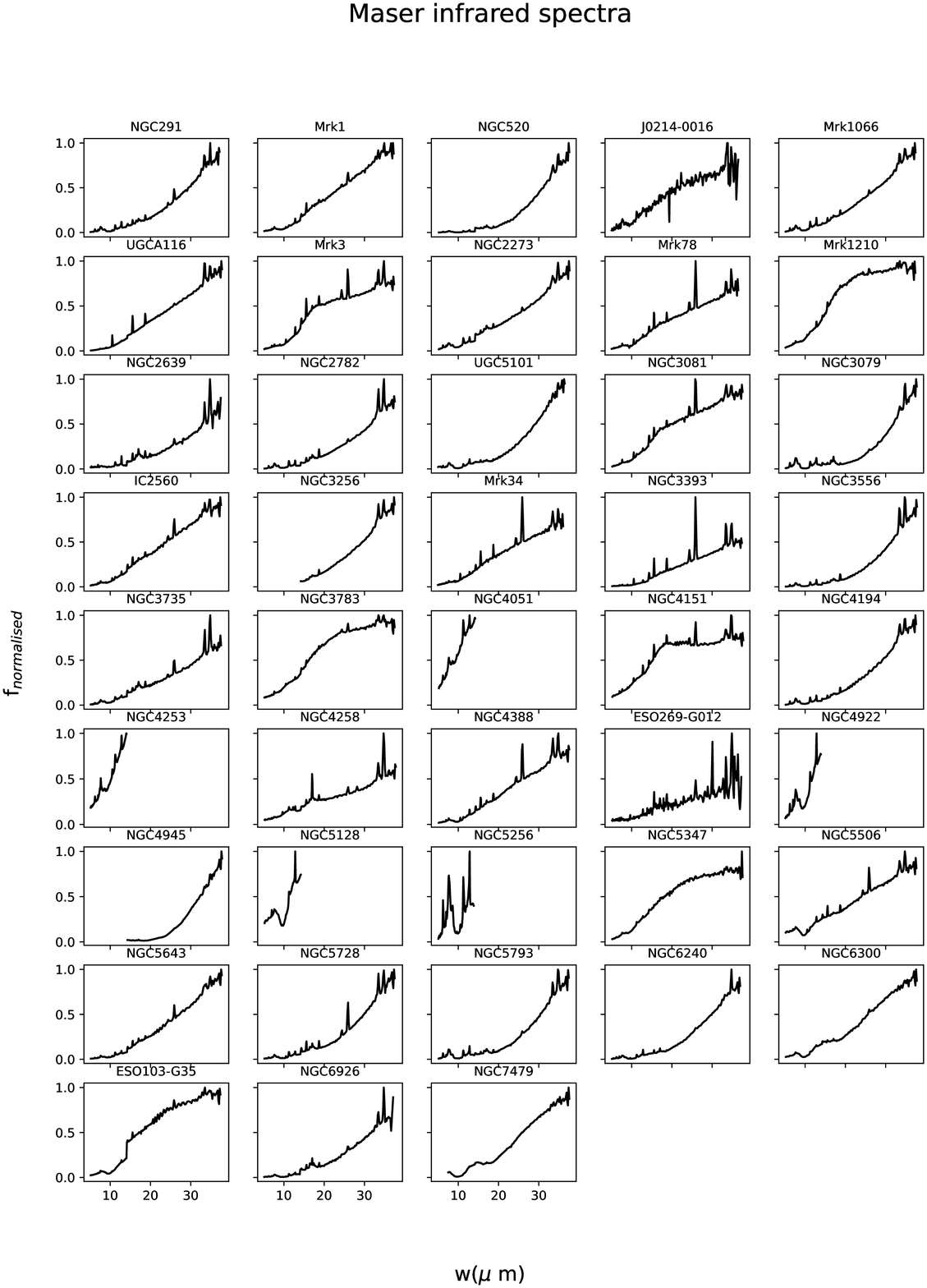}
    \caption{The mid-infrared spectra of all selected maser galaxies.}
    \label{fig:spec_masers_all}
\end{figure*}

All mid-infrared spectra of selected masers are listed in the Figure~\ref{fig:spec_masers_all}.

\bsp	
\label{lastpage}
\end{document}